\documentclass[twocolumn]{aastex631}
\usepackage{enumitem}
\usepackage{amsmath}
\usepackage{float}
\usepackage{natbib}
\usepackage{cancel}
\usepackage{xcolor,soul}
\usepackage[normalem]{ulem}
\useunder{\uline}{\ul}{}
\usepackage{hyperref}
\usepackage{bold-extra}
\usepackage{bm}
\usepackage{ragged2e}

\newcommand{\hh}{\text{H}_2}
\newcommand{\fhh}{f_{\text{H}_2}}
\newcommand{\mhh}{M_{\text{H}_2}}
\newcommand{\fgas}{f_{\text{gas}}}
\newcommand{\mgas}{M_{\text{gas}}}
\newcommand{\lco}{L'_{\text{CO}}}
\newcommand{\sfr}{\text{SFR}}
\newcommand{\ssfr}{\text{sSFR}}
\newcommand{\tH}{\text{t}_\text{H}}
\newcommand{\eq}{Eq.\hspace{.2em}}
\renewcommand{\fig}{Fig.\hspace{.2em}}

\newcommand{\app}{Appendix\hspace{.2em}}
\newcommand{\mbh}{M_{\text{BH}}}

\newcommand{\fedd}{f_{\text{Edd}}}
\newcommand{\simba}[0]{{\sc simba} }

\newcommand{\yt}[0]{{\sc yt} }
\newcommand{\salad}[0]{{\sc caesar} }
\newcommand{\powderday}[0]{{\sc powderday} }
\newcommand{\hyperion}[0]{{\sc hyperion} }
\newcommand{\fsps}[0]{{\sc fsps} }
\newcommand{\slick}[0]{{\sc slick} }
\newcommand{\gizmo}[0]{{\sc gizmo} }
\newcommand{\eagle}[0]{{\sc eagle} }
\newcommand{\tng}[0]{{\sc IllustrisTNG} }

\newcommand{\despotic}[0]{{\sc despotic} }
\newcommand{\magnet}[0]{{\sc Magneticum Pathfinder}}
\newcommand{\aco}{\alpha_\text{CO}}
\newcommand{\acomw}{\alpha_{\text{CO,MW}}}

\newcommand{\colibre}[0]{{\sc colibre}}
\newcommand{\squig}[0]{SQuiGG$\vec{L}$E}
\graphicspath{{./}{Figures/}}

\shorttitle{The Mystery of The Red, Dead \& Gassy Galaxies}
\shortauthors{Savitch et al.}
\begin{document}
\title{Unveiling the Origins of the Molecular Gas Reservoirs in Recently Quenched Massive Galaxies}

\author[0000-0002-2919-1109]{Ethan Savitch \href{mailto:esavitch@ufl.edu}{(esavitch@ufl.edu)}}
\affiliation{University of Florida Astronomy, 211 Bryant Space Science Center, Gainesville, FL, 32611, USA}

\author[0000-0002-7064-4309]{Desika Narayanan}
\affiliation{University of Florida Astronomy, 211 Bryant Space Science Center, Gainesville, FL, 32611, USA}
 \affiliation{Cosmic Dawn Center (DAWN), Copenhagen, Denmark}

\author[0000-0003-4075-7393]{David Setton}
\affiliation{Department of Physics and Astronomy and PITT PACC, University of Pittsburgh, Pittsburgh, PA 15260, USA}

\author[0000-0001-7160-3632]{Katherine E. Whitaker}
\affiliation{Department of Astronomy, University of Massachusetts, Amherst, MA 01003, USA}
\affiliation{Cosmic Dawn Center (DAWN), Copenhagen, Denmark}

\author[0000-0001-5063-8254]{Rachel Bezanson}
\affiliation{Department of Physics and Astronomy and PITT PACC, University of Pittsburgh, Pittsburgh, PA 15260, USA}

\author[0000-0002-0766-1704]{Elia Cenci}
\affiliation{Department of Astronomy, University of Geneva, Chemin Pegasi 51, Versoix CH-1290, Switzerland}
\affiliation{Department of Astrophysics, Universit\"at Z\"urich, Winterthurerstrasse 190, Zurich CH-8057, Switzerland}

\author[0000-0003-2842-9434]{Romeel Dav\'e}
\affiliation{University of Edinburgh, Institute for Astronomy}
\affiliation{Department of Physics and Astronomy, University of the Western Cape, Bellville, Cape Town 7535, South Africa}

\author[0000-0002-1759-6205]{Vincenzo R. D'Onofrio}
\affiliation{Department of Physics and Astronomy and George P. and Cynthia Woods Mitchell Institute for Fundamental Physics and Astronomy, Texas A\&M University, 4242 TAMU, College Station, TX 77843-4242, USA}

\author[0000-0002-1109-1919]{Robert Feldmann}
\affiliation{Department of Astrophysics, Universit\"at Z\"urich, Winterthurerstrasse 190, Zurich CH-8057, Switzerland}

\author[0000-0002-5612-3427]{Jenny Greene}
\affiliation{Department of Astrophysical Sciences, Princeton University, 4 Ivy Lane, Princeton, NJ 08544, USA}

\author[0000-0002-7613-9872]{Mariska Kriek}
\affiliation{Leiden Observatory, Leiden University, P.O. Box 9513, 2300 RA Leiden, The Netherlands}

\author[0000-0002-9729-3721]{Massimiliano Parente}
\affiliation{University of Florida Astronomy, 211 Bryant Space Science Center, Gainesville, FL, 32611, USA}

\author[0000-0003-3256-5615]{Justin Spilker}
\affiliation{Department of Physics and Astronomy and George P. and Cynthia Woods Mitchell Institute for Fundamental Physics and Astronomy, Texas A\&M University, 4242 TAMU, College Station, TX 77843-4242, USA}

\author[0000-0002-1714-1905]{Katherine A. Suess}
\affiliation{Department for Astrophysical \& Planetary Science, University of Colorado, Boulder, CO 80309, USA}

\author[0000-0003-0372-5253]{Dariannette Valent\'in-Mart\'inez}
\affiliation{University of Florida Astronomy, 211 Bryant Space Science Center, Gainesville, FL, 32611, USA}

\author[0009-0008-7017-5742]{Dhruv Zimmerman}
\affiliation{University of Florida Astronomy, 211 Bryant Space Science Center, Gainesville, FL, 32611, USA}

\begin{abstract}
Observations of quenched galaxies across cosmic time have revealed systems containing substantial molecular gas reservoirs, challenging the traditional picture of gas-poor, red and dead galaxies. Using the \simba cosmological simulation, we investigate three possible hypotheses for the origin of molecular gas-rich quenched systems: (i) quiescence is temporary and the galaxy will rejuvenate, (ii) molecular gas masses are overestimated due to an underestimated CO-to-H$_2$ conversion factor, or (iii) star formation is ongoing but heavily obscured by dust. Across $M_*>3\times10^9M_\odot$ galaxies from $z=0-2$, we find that all three mechanisms are plausible. Many seemingly quiescent, molecular gas-rich galaxies host dust-obscured star formation, with visual optical depths increasing on average from $\tau_V\sim0.5-2.3$ from $z=0-2$. Assuming a Milky Way-like CO-to-H$_2$ conversion factor can overestimate molecular gas masses by a factor of $\sim\!5$ in non-star-forming galaxies, compared to $\sim\!1.45$ in star-forming galaxies.  Temporary quiescence can also occur in gas-rich transitioning galaxies on timescales ranging from $100\,\text{Myr}-2\,\text{Gyr}$, although this is most prominent at high redshift and is relatively rare overall.  Further work is needed to disentangle the relative contributions of these mechanisms across cosmic time.
\end{abstract}

\section{Introduction}
\label{sec:intro}
Observations of galaxies in the local universe find a persistent bimodality in galaxy populations \citep[e.g.][]{noeske,wuyts,Brammer11, Muzzin_2013,jin}.  This dichotomy is defined by one population of star-forming galaxies that are blue in color with spiral morphology \citep{saintonge, tacconi}, and a second population, traditionally called quiescent or `red and dead' galaxies, which have their star formation quenched and have typically early-type morphological structures \citep{blanton03, mcdermid}.  In the Local Universe, star-forming galaxies are observed to be gas-rich relative to their quiescent counterparts \citep{davis,young}.

Studies of massive galaxies at high-$z$ confirm that the majority of quenched systems form their stars early on and in short bursts \citep{carnall, Belli,merlin, suess21, laporte, Tacchella_2022}.  One such product of this rapid quenching are the so-called ``post-starburst galaxies" (hereafter `PSB' galaxies), which are thought to be the direct result of massive galaxies undergoing a strong burst of star formation that dramatically shuts down within $\lesssim$ 1 Gyr \citep{dressler, zabludoff, goto05, goto07, yang, kriek10, wild20, french_2021}.  PSB galaxies are most common at Cosmic Noon, contributing to half of the total growth of quenched galaxies at $z\sim 2$ \citep{Belli}, though they are relatively rare in the local universe, representing only $\sim 20\%$ of transitioning galaxies at $z\sim 1.4$ and a negligible fraction at $z\sim 0$ \citep{tran,whitaker12,wild16, Lotz21,Setton_2023, smercina, park24}. 

Since such galaxies are assumed to be shutting down their star formation and transitioning to being fully quenched, traditional star formation laws suggest that they should be relatively devoid of molecular gas \citep[e.g.][]{Kennicut, shmidt}.  In recent years, there has been a flurry of activity in observationally studying the cold gas reservoirs in PSB galaxies \citep{rowlands, french_2015, Alatalo, alatalo17,suessSquiggle, setton2025}. Contrary to traditional expectations, such observations have revealed that PSB galaxies can retain significant reservoirs of CO-traced molecular (H$_2$) gas, with molecular gas fractions up to $\sim 20\%$ at $z<1$, even after shutting down their star formation \citep{young, suess_2017, French_2018, Bezanson_2022, Baron22, setton2025, Umehata2025}.  Interestingly, the molecular gas fraction in these galaxies is a strong function of the time elapsed since quenching \citep{suess2025}.  However, as noted by \cite{Bezanson_2022}, the inferred gas depletion timescales for such galaxies ($\sim$Gyr) is significantly longer than their observed quenching timescale ($\sim$100-200Myr).  These findings indicate that the quenching of star formation in PSBs is not directly driven by gas exhaustion; instead suggesting that the molecular gas reservoirs are rendered inhospitable to star formation either through the heating or removal of ISM gas.

While quenching in such systems has been studied since \citet{dressler}, a clear picture of the processes driving its onset has yet to emerge.  The most commonly proposed mechanisms include processes internal to galaxies (morphological quenching or AGN feedback;\,e.g.,\,\citealt{Martig09, Baron18, Calabro19}) as well as environmental effects like ram-pressure stripping, galaxy mergers, and starvation (e.g.,\! \citealt{blanton09,snyder, Lotz21, Spilker22, Donofrio25}).  However, the timescales associated with many of these mechanisms are not able to account for the rapid quenching observed at high redshift \citep{Yesuf14, french_2017, winds, french_2021, Bezanson_2022}. These questions, along with the difficulty of explaining the large molecular gas abundances detected in such systems \citep[e.g.,][]{YusefHo20, cenci25, setton2025}, pose a challenge for models of galaxy evolution.

Motivated by this, the goal of this paper is to holistically investigate the various proposed pathways that can result in recently quenched high-mass galaxies preserving their molecular gas with cosmological galaxy simulations.  In particular, we investigate three potential physical scenarios for the existence of copious molecular gas reservoirs remaining in massive galaxies with seemingly low star-formation rates.

Our first hypothesis is that recently quenched massive galaxies, with observable levels of molecular gas, may simply be in a temporary period of dormancy, and are on a path to rejuvenation \citep{remus2023,Kimmig25}.  Multiple theoretical studies support this: \cite{Lotz21} found that $\sim23\%$ of massive quiescent galaxies undergo rejuvenation when tracked forward from $z=3.4$ in the \magnet\ simulation suite. \cite{harrold2025} used a semi-analytic model to show that a typical massive and recently quenched galaxy at $z=1$ has a $\sim25\%$ chance of rejuvenating within 1 Gyr.  Finally, \cite{colibre} used the new \colibre\ simulations to show that up to half of massive quenched galaxies could rejuvenate after initial quenching.  Observational studies independently support this hypothesis, such as the work by \cite{ellison25} in which FAST 21 cm observations of 68 PSB galaxies were used to show that at low redshift these systems can contain HI gas fractions up to 30\%.  The presence of substantial atomic gas reservoirs observed in these rapidly quenched galaxies suggests that such PSB-like systems do not quench through the wholesale removal of their gas, and therefore remain viable candidates for future rejuvenation.

The remaining two hypotheses that we investigate deal with observational effects that could result in the inference of molecular gas present in seemingly quenched galaxies.  Possible solutions could be that the molecular gas is not as abundant as it seems to be (Hypothesis 2), or that the galaxy is not actually as quenched as it appears (Hypothesis 3).  In particular, our second hypothesis investigates whether molecular gas mass estimates are impacted by potentially varying CO-to-H$_2$ conversion factors \citep{narayanan12, Bolatto_2013}, and our third hypothesis asks whether dust attenuation mimics colors compatible with selection as a quenched galaxy \citep{uvj, Baron2023, psbEagle, cenci25}.  The latter hypothesis is also supported by evidence that typical massive quenched galaxies at cosmic noon ($z\sim 2.5$) are significantly more reddened than those in the local universe, with dust attenuation falling from $A_V\sim 0.5$ at $z=2.5$ to near zero at present times \citep{siegel24}. 

This paper is organized as follows.  In \S\ref{sec:methods}, we describe the numerical methods that we employ, starting with the simulation we use (\S\ref{ssec:simba}), the dust model it implements (\S\ref{sssec:dust}), and the updates to its black hole feedback model (\S\ref{ssec:bh}).  In \S\ref{ssec:prog}, we discuss how we track galaxies across different snapshots using progenitor and descendant tracking.  In \S\ref{ssec:selectGal} we define our galaxy selection criteria, first by defining our method for classifying quenched/transitioning galaxies (\S\ref{ssec:select}), followed by an explanation of the photometric selection criteria we use to investigate observational systematics (\S\ref{ssec:ubvSelect}).  In \S\ref{sec:results}, we present the results of our analysis: specifically, in \S\ref{ssec:HA} we examine temporary quiescence, in \S\ref{ssec:HD} the impact of $\aco$, and finally in \S\ref{ssec:HC} we investigate the impact of dust obscuration.  Next, in \S\ref{sec:discussion}, we discuss how our simulation and methodology compares to other analogous numerical studies.  Finally, in \S\ref{sec:conclusion} we conclude by summarizing our results, their implications, and avenues for future work on this topic.

\section{Methods}
\label{sec:methods}
\subsection{Cosmological Simulations}
\label{ssec:simba}

We employ the \simba galaxy formation simulation \citep{simba}, which is a modern cosmological simulation that includes models for star formation, its associated feedback, black hole growth/feedback, and interstellar dust.  Specifically, we employ the fiducial $(25/$h Mpc)$^3$ box with $2\times512^3$ particles, resulting in a baryon mass resolution of $M_{\rm b} \approx 2\times10^6 M_\odot.$   

The \simba simulation employs the meshless finite mass (MFM) hydrodynamic solver within the \gizmo codebase \citep{hopkins15,hopkins17}.  \simba includes a sub-resolution prescription for star formation and feedback.  Here, star formation occurs stochastically in H$_2$ gas, where the molecular content is determined using the \citet*{KMT07} algorithm, which calculates the fraction of gas in molecular hydrogen ($f_{\hh}\equiv \mhh/\mgas$) as a function of metallicity and column density \citep{mufasa}.  More information about this algorithm can be found in \cite{KMT11}.

Star formation follows a volumetric \citet{shmidt} law (hereafter SFR $\equiv\dot{M}_*=$ Star Formation Rate): 
\begin{equation}
\sfr\equiv \epsilon_*\bigg(\frac{\rho_{\rm gas}\times  \fhh}{t_{\rm ff}}\bigg)\propto \rho_{\rm gas}^{3/2}
\label{eq:sfr}
\end{equation}
where $\epsilon_*=0.02$ is the fixed efficiency of star-formation per free-fall time, and $t_{\rm ff}$ is the free-fall time-scale ($t_{\rm ff}=\sqrt{3\pi/(32G\rho_{\rm gas})} \propto \rho_{\rm gas}^{-1/2}$).  By construction, the star-formation rate is directly dependent on the density of its gas ($\rho_{\rm gas}$), and its molecular gas fraction ($\fhh$).  In their formalism, they also require a minimum density for star formation of $n_H\geq 0.13\, \text{cm}^{-3}$, defining gas with densities above this as `ISM gas' \citep{mufasa}.

\subsubsection{Dust Model}
\label{sssec:dust}
\simba includes a model for dust physics that is able to capture the formation and evolution of cosmic dust.  Due to the inability to resolve gas drag in our simulation, along with the very short time scales associated with it, the model has dust passively follow the flow of gas particles. The modeled dust grains have a single fixed radius of $a=0.1~\mu \text{m}$, and active dust cooling is ignored \citep{simba}.

Following the prescription of \cite{dwek}, cosmic dust is produced by metal condensation in the ejecta of supernovae (SNe) and Asymptotic Giant Branch (AGB) stars.  After being produced, the model allows dust to grow in mass by accreting gas-phase metals; in which dust particles collide with gaseous metals, causing the metals to stick to the dust grains and allowing their mass to grow.  In addition, thermal sputtering is implemented so that dust grains can be eroded by thermally excited gas  \citep{simba}.  Finally, since the simulation cannot resolve individual supernovae (SNe) shocks, an additional dust destruction mechanism by SNe shocks is included, following the prescription from \cite{mckinnon}.  The dust model, fully detailed in \cite{Li2019}, shows good agreement with observational constraints on dust abundance, especially its redshift evolution \citep{parente2025}.

\subsubsection{Black Hole Model}
\label{ssec:bh}
\simba includes on-the-fly black hole growth and AGN feedback, fully described in \cite{simba}; here we summarize the features salient to this work.  Black holes grow via a two-mode accretion model: torque-limited accretion from cold gas ($T<10^5\,\text{K}$), following \citet{Hopkins2011} and \cite{Angles2017}, and \cite{Bondi1952} accretion from hot gas ($T>10^5\,\text{K}$).  Feedback is implemented kinetically as bipolar outflows, with the mode set by the Eddington ratio of the galaxy's most massive black hole.  At $\fedd>0.2$, AGN drive multiphase `radiative mode' winds, which have only a modest effect on galaxy populations \citep{dave20}.  When $\fedd<0.2$ and $\mbh>10^{7.5}M_\odot$, feedback transitions to `jet mode', driving hot collimated outflows whose velocity increases as $\fedd$ drops, up to an additional 7000 km s$^{-1}$ at $\fedd\leq0.02$.  Finally, when $\fedd<0.02$ and $\fgas\equiv \mgas/M_{*}<0.2$, X-ray heating of the ISM is included, imparting a radial outward kick and temperature increase.

The jet mode is primarily responsible for quenching and maintaining the massive quenched population in \simba, heating the circumgalactic gas and shutting off the inflow of fresh gas \citep{simba, Hardcastle07}, while the X-ray mode supplies the additional energy needed to quench star formation in massive galaxies.

\subsection{Progenitor/Descendant Tracking}
\label{ssec:prog}

After the execution of the \simba simulation, we employ \salad \citep{salad}, a \yt\!\!-based package \citep{yt}, to identify dark-matter halos and galaxies using a 6D friends-of-friends algorithm.  For each snapshot, \salad outputs a cross-matched halo and galaxy HDF5 catalog, with an abundance of precomputed properties for each \citep{simba}, as well as the particle data used to calculate them.  In order to track a single galaxy through time, we must link galaxies across multiple snapshots: here we describe this progenitor and descendant tracking of galaxies through cosmic time.

In order to track a single galaxy backwards through time (to higher $z$) we must use progenitor tracking.  Since any one galaxy would have had many ancestral galaxies that merged to produce it, tracking galaxies backwards in time requires following \textit{all} galaxies that share common star particles with it at any point in the past.  In practice, we identify a galaxy of interest at $z=0$, and then increment through successive higher redshift snapshots, searching for all galaxies that share at least $10\%$ star particles in common with our galaxy of interest, all of which will be considered its progenitors.  Since all such galaxies will eventually contribute to our target galaxy at $z=0$, we must sum up \textit{all} of their properties to effectively compare across snapshots.  Conceptually, this is essentially calculating the merger tree for any target galaxy.

While this method works for moving backward in time, if we wanted to track a galaxy forward in time (to lower $z$) we would instead be tracking its descendants.  This becomes necessary, for example, when we want to see at a given redshift if a galaxy will rejuvenate as time increases.  As discussed in the literature \citep[e.g.][]{Torrey_2015}, this process is fundamentally different from progenitor tracking due to the nature of mergers.  While tracking a galaxy backwards in time using a merger tree requires calculating the sum of all the progenitors, in our formalism tracking it forward in time requires only the most massive parent node to be tracked up the tree.  This is because while tracking galaxies forward in time a significant fraction of the population can be `lost' via merger events, no such loss occurs when tracking backward in time \citep{Torrey_2015}.

\subsection{Galaxy Selection Criteria}
\label{ssec:selectGal}
\label{ssec:massLimit}
\begin{figure*}[htb!]
    \centering
    \includegraphics[width=2\columnwidth]{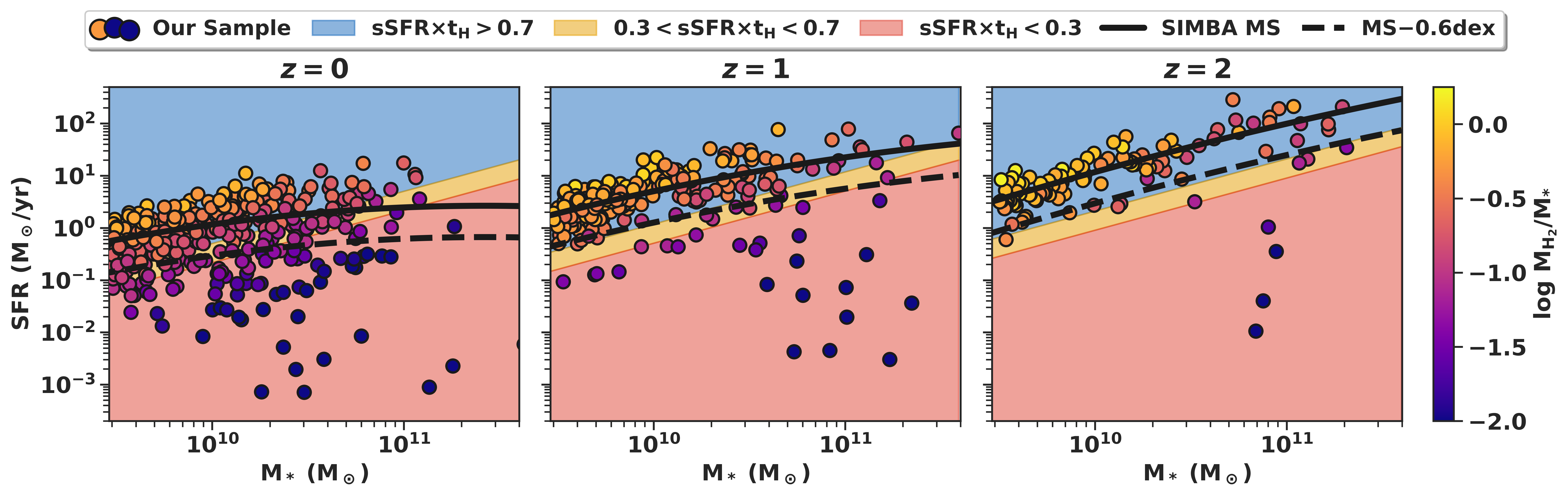}
      \caption{{\bf Full galaxy sample shown along with our adopted selection criteria, as well as with a comparison to alternate methods.}  Each point represents a $M_*>3\times 10^9M_\odot$ galaxy in our \simba simulation, with a color corresponding to the log ratio of total molecular $(\hh)$ to stellar mass.  The blue shaded region highlights where we define galaxies to be `star forming' (\eq\ref{eq:ssfrt1}), while the yellow region below that corresponds to `transitioning' (\eq\ref{eq:ssfrt2}), and the red region below that for fully `quenched' (\eq\ref{eq:ssfrt3}).  For comparison, the star forming main sequence (SFMS) for {\scshape simba} is shown as solid black line \citep[derived by fitting our data to the functional form of][\eq \ref{eq:sfms}]{sfms}, while the dashed black line shows where galaxies have SFRs that lie 0.6 dex below the SFMS.  While in general our definition of quenched galaxies (\eq\,\ref{eq:ssfrt3}) captures approximately the same galaxy population as are 1 dex below the SFMS, we see that at $z=0$ the two definitions differ most at the high-mass end.  A full comparison of these methodological choices is explored in \app\,\ref{app:ssfrCompare}}
    \label{fig:sfms}
\end{figure*}

We focus our study on massive quenched galaxies comparable to current theoretical work \citep{psbEagle,setton2025} and observational surveys \citep[e.g.][]{rowlands, french_2015, french_2017, French_2018, suessSquiggle, Bezanson_2022, Baron2023, french23, Spilker2025}, and therefore impose a minimum mass requirement.  Following the methodology of \cite{cenci25}, in order to compare best with recent observations such as \cite{suessSquiggle} we restrict our analysis to galaxies with stellar masses $M_*>3\times 10^{9}M_\odot$, producing an initial sample of 395 galaxies at $z=0$ (245 at $z=1$, 110 at $z=2$, and 60 at $z=3$).  All galaxies satisfying this mass cut are shown on a plot of stellar mass vs SFR in \fig \ref{fig:sfms} for $z=0,1,2$.

\subsubsection{Star Formation Classification}
\label{ssec:select}
Following \cite{mergers,Tacchella_2022}, we employ a combination of the specific star formation rate ($\ssfr\equiv \sfr/\text{M}_{*}$) and the age of the Universe at the time of galaxy identification ($\tH\equiv 1/\text{H(z)}$) to decide whether a galaxy is star-forming ($\ssfr\geq 0.7/\tH$), fully quenched ($\ssfr\leq 0.3/\tH$), or intermediate/transitioning between the two:
\begin{align}
    \textrm{Star Forming}:\quad &\ssfr\times \tH \geq 0.7\label{eq:ssfrt1}\\
    \textrm{Transitioning}:\quad &0.7>\ssfr\times \tH>0.3\label{eq:ssfrt2}\\
    \textrm{Quenched}:\quad &\ssfr \times \tH \leq 0.3\label{eq:ssfrt3}
\end{align}
Specifically, since $1/\ssfr$ gives the timescale for a galaxy to double its stellar mass, as described in \cite{Tacchella_2022}, comparing it to $\tH$ gives an estimate of the number of times the stellar mass has doubled within the age of the universe (assuming a constant $\ssfr$).  To put this into context, the Milky Way would have $\ssfr\times\tH\sim0.4$ \citep{Licquia15}, thus would be defined as `transitioning', have a stellar mass-growth timescale of $2.5\tH$, and at its current level of star-formation would have formed 40\% of its existing stellar mass over a Hubble time.

In \fig\ref{fig:sfms}, we show the regime we would define as `star-forming' according to \eq\ref{eq:ssfrt1} with a blue background, `transitioning' (\eq\ref{eq:ssfrt2}) in yellow, and `quenched' (\eq\ref{eq:ssfrt3}) in red.  While a quenching threshold of $\ssfr\times \tH\leq 0.2$ is more common \citep{Tacchella15, Pacifici16, whitakerReview}, we adopt the slightly more inclusive value of 0.3 \citep[following, e.g.,][]{Franx08, Williams10, Szomoru13, Lotz19}, as our aim is to broadly capture low star-forming systems rather than isolate the most strictly quenched galaxies.

Among our initial sample of $M_*>3\times 10^{9}M_\odot$ galaxies, we find that at $z=0$ that 249 (63\%) are ``star-forming", 54 (14\%) are ``transitioning", and 92 (23\%) are fully ``quenched".  This is consistent with the numerical results from \cite{Donnari19}, which used rest-frame UVJ photometric colors to identify star-forming galaxies in the \tng simulations, predicting that 71\% of $M_*>10^{9}M_\odot$ galaxies will be star-forming.  Moving to higher redshift, we find that at $z=1$ the number of ``star-forming"/``transitioning"/``quenched" galaxies decreases to 213/7/25 (87\%/3\%/10\%), at $z=2$ this becomes 99/6/5 (90\%/5\%/5\%), and at $z=3$ we find 59/0/1 (98\%/0\%/2\%).  While the lack of high-mass galaxies at high-$z$ is what causes the numbers to decrease so rapidly, we find in contrast that the fraction of star-forming galaxies increases with redshift.  This is consistent with a number of observational studies, such as \cite{Weaver23}, who studied galaxies from $z=7.5\to2$ from the COSMOS2020 observational catalog using their rest-frame near-ultraviolet (NUV), r, and j band colors, and found that at $z\geq2$, around $90\%$ of galaxies will be star-forming.

An alternative method for classifying star-forming galaxies is to measure their offset from the Star-Forming Main Sequence (SFMS; e.g. \citealt{noeske, whitaker12, sfms,Speagle14}), the well-established relation between stellar mass ($\text{M}_{*}$) vs star-formation rate ($\sfr$).  Since galaxies become increasingly quiescent as they move below the SFMS, systems lying more than 0.6 dex beneath the relation are considered `transitoning', while those more than 0.9 dex below are fully `quenched' \citep{Donnari19, Bluck20, whitakerReview}. 

To see how this selection method compares to our own, we show in \fig\ref{fig:sfms} the SFMS for \simba a solid black line, calculated by fitting our data at each redshift to the functional form of \cite{whitaker12} (via \eq \ref{eq:sfms}), with details provided in \app \ref{app:ssfrCompare}, and an associated dashed black line showing where galaxies are 0.6 dex below this SFMS.  In our formalism (\eq\ref{eq:ssfrt1}-\ref{eq:ssfrt3}), this dashed line can be compared to the top of our yellow region, which we use to show where our transitioning galaxies lie (\eq\ref{eq:ssfrt2}).  Doing so shows that both definitions capture a similar population of non star-forming galaxies (i.e. transitioning or quenched), with the worst correspondence occurring at low-$z$ and high mass.  This arises from the known fact that the SFMS slope flattens at the massive end at low redshift \citep{Abramson2014}, possibly due to the combination of an older bulge and younger disk \citep{Whitaker15}. 

In \app\,\ref{app:ssfrCompare} we describe in detail how we calculated the SFMS for \simba\!\!, how it is traditionally used to categorize star-forming galaxies, and how this definition differs from the sSFR$\times\tH$ definition we used in this study.  In summary, given these differences, we choose to avoid using a SFMS-based definition for quenching, partly because the redshift evolution of the SFMS differs between simulations and observations \citep{bluck2023galaxy}, and also as shown in \app\,\ref{app:ssfrCompare}, our sSFR-based method of identifying `low star-forming' systems produces a population of systems much more gas-rich than if we used the SFMS method.  Therefore, we adopt our sSFR-based definition of quenching, given in \eq \ref{eq:ssfrt1}-\ref{eq:ssfrt3}, as it is non-dimensional, invariant across redshift, and yields a more gas-rich sample.

\subsubsection{UBV-Selection}
\label{ssec:ubvSelect}
While for the majority of the analysis we want to use a physical definition of quenching, as described in the previous section (\S\ref{ssec:select}), there will be a point where we will investigate potential observational systematics in selecting quenched systems.  Specifically, this becomes important while investigating our third hypothesis (\S\ref{ssec:HC}), in which we ask whether dust could obscure ongoing star formation in seemingly quenched systems, creating false signatures of quenching in otherwise star-forming, molecular gas rich galaxies.  Because of the nature of this hypothesis, we naturally want to select galaxies in as faithful to an observational manner as possible.

To derive mock colors for our simulated galaxies, we must first create synthetic spectra.  To generate realistic mock spectral energy distributions (SEDs) for our sample of galaxies, we post-process our simulated snapshots using the dust radiative transfer code \powderday (\url{https://github.com/powderday-rt/powderday}).  First introduced in \cite{powderday}, \powderday uses \fsps for stellar population synthesis models \citep{conroy1, conroy2, conroy3}, \hyperion for radiative transfer \citep{hyperion}, and \yt to interface between varying software packages \citep{yt}.  

When running \powderday to produce mock SEDs, we smooth the particle data onto an octree to setup the model for radiative transfer.  Next, we employ \fsps to generate a simple stellar population (SSP) for each star particle.  For consistency with \simba\!\!\!, we use a Chabrier initial mass function (IMF) \citep{simba}.  Combined with MIST isochrones \citep{Choi16, Dotter_2016} and a MILES stellar spectral library \citep{Sanchez06} as our fiducial SPS parameters, we then run \fsps to create a dust-free stellar SED.  Given the initial stellar SED from \fsps and the grid from \yt\!\!, \hyperion is then employed to run 3D radiative transfer via a Monte Carlo algorithm to produce the final SED \citep{hyperion}.  Finally, the resulting SED is extracted by assuming the largest aperture size available to the model.

To construct our photometric selection criteria, we follow the \cite{kriek10} method, also used by \cite{suessSquiggle}, which uses synthetic medium-band rest-frame filters ($U_m$, $B_m$, and $V_m$) to define ``post-starburst galaxies" (hereafter PSBs) as 
\begin{equation}
\begin{aligned}
\underline{\text{UBV-Selected}}& \text{ (``PSB-like") Galaxies:}\\
U_m-B_m>0.975\ \ \ &\&\ \ -0.25<B_m-V_m<0.45
\end{aligned}
\label{eq:UBVSelect}
\end{equation}

According to \cite{suessSquiggle}, the $U_m-B_m$ color is able to separate PSBs from unobscured star-forming galaxies by quantifying the strength of the Balmer break, while the $B_m-V_m$ color is able to separate PSBs from dusty star-forming galaxies and older quiescent galaxies by quantifying the slope of the spectra redwards of the break. Collectively, these photometric cuts are meant to identify relatively dust-free galaxies that recently quenched after a large starburst.  Since we are using this method to identify our sample of galaxies in \S\ref{ssec:HC}, it is informative to check how our resulting sample differs from the sample produced by \cite{suessSquiggle}, which is done in \app \ref{app:ubv}.

Applying this photometric cut to our initial sample of $M_*>3\times 10^{9}M_\odot$ galaxies results in a sample of 21 (5.3\%) UBV-selected galaxies at $z=0$. In agreement with the findings from \cite{Belli} that PSB galaxies are most common at Cosmic Noon, we find that the highest fraction of UBV-selected galaxies in our sample occurs at $z=2$, dropping off both at higher and lower redshifts than this.  Specifically, at $z=1$ we find 48 galaxies (20\%) in our sample are UBV-selected, at $z=2$ we find 30 (27\%), while at $z=3$ this rapidly drops to only 10 galaxies (17\%) which satisfy our UBV-selection. 

\begin{figure}
    \centering
    \includegraphics[width=\linewidth]{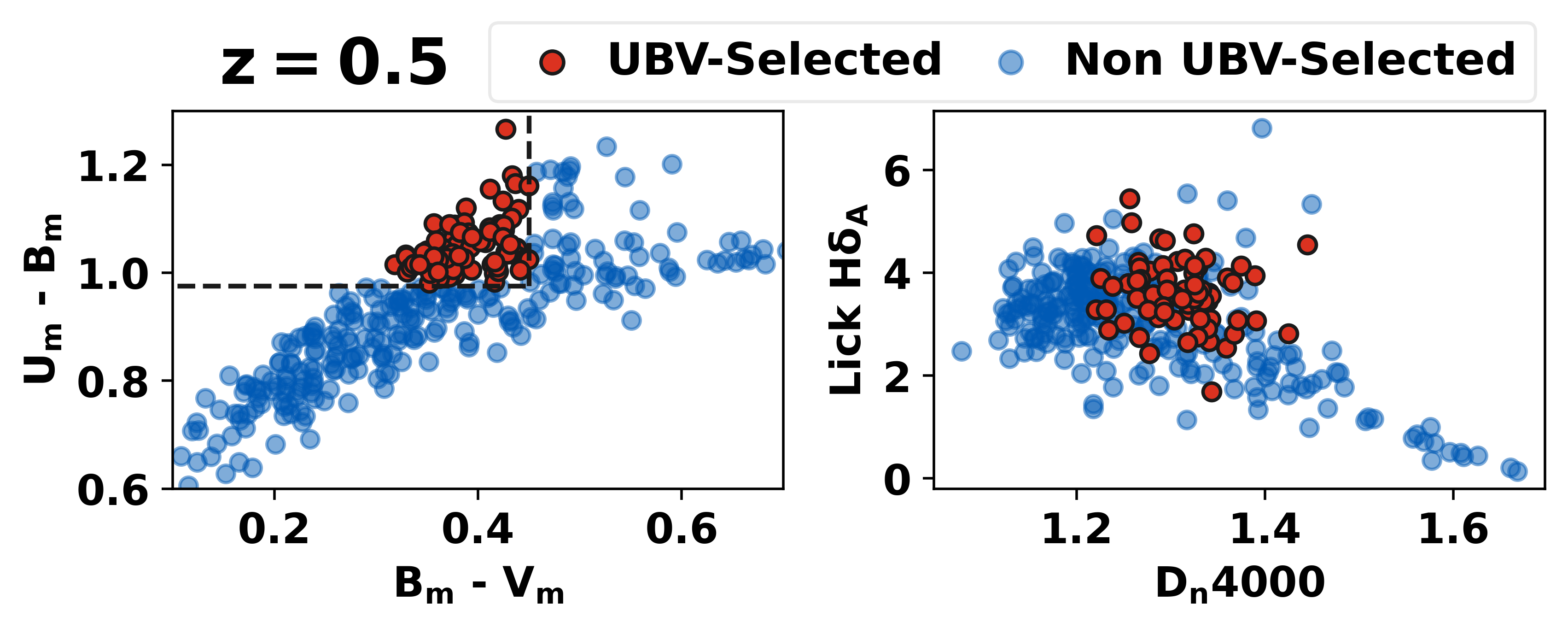}
    \includegraphics[width=\linewidth]{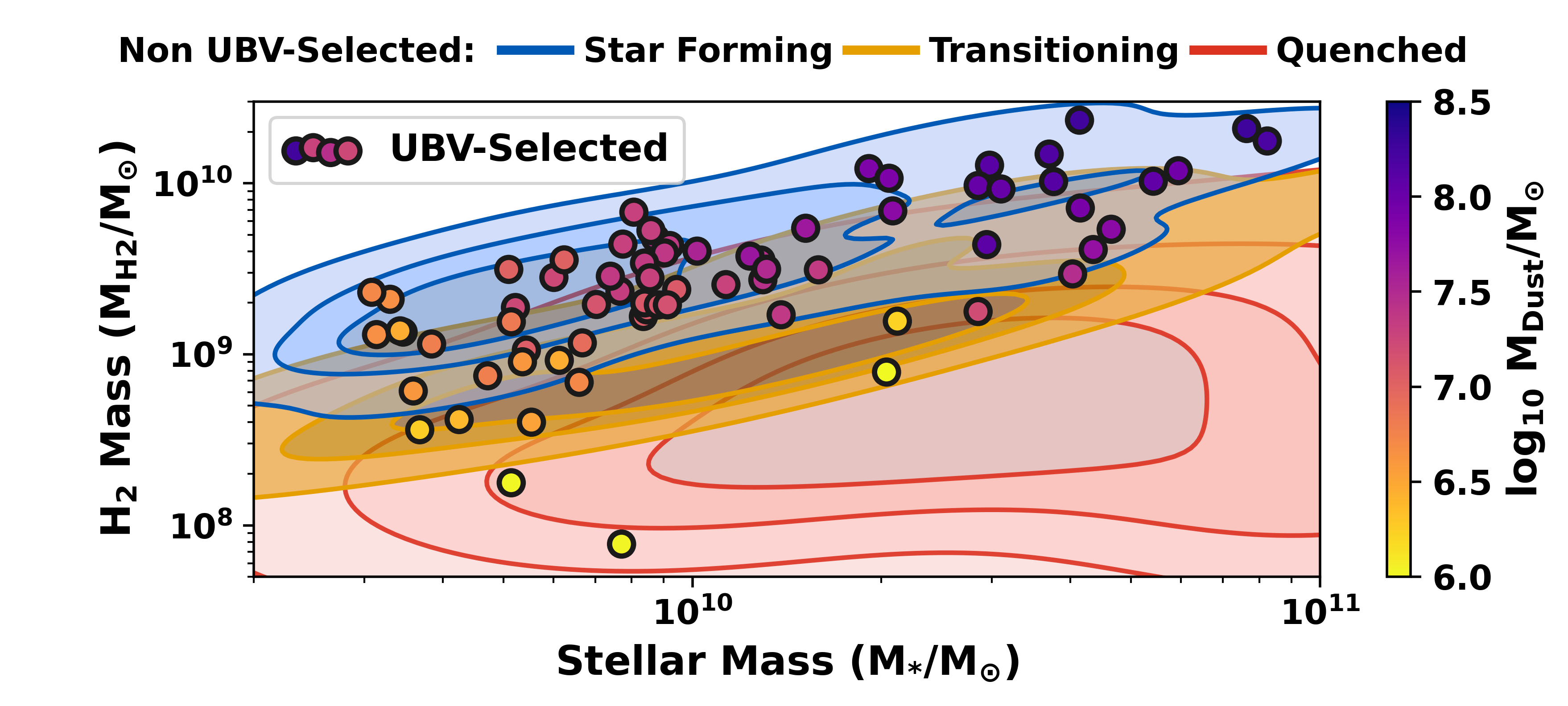}
    \caption{{\bf (\underline{Top}) The distribution of our {\bfseries \scshape simba} galaxies at $\bm{z=0.5}$ in UBV space (left), and in a plot of D$\bm{_n}$4000 vs H$\bm{\delta_A}$ (right).}  The red-colored points are those that satisfy the PSB-like UBV selection (\eq\ref{eq:UBVSelect}) from \cite{suessSquiggle}, defined to identify relatively dust-free galaxies that recently quenched after a large starburst.  The left-hand plot shows the galaxies in UBV-space, defined by the medium-band rest-frame filters ($U_m$, $B_m$, and $V_m)$, designed to target the Balmer break and slope of the spectrum redward of the break \citep{suessSquiggle}. Right-hand plot shows the galaxies lick H$\delta_A$ index (i.e. H$\delta$ EW; which traces recent star formation) as a function of the 4000\r{A} break index (which probes the age of the stellar population), such that age increases toward the lower right. {\bf(\underline{Bottom}) UBV-selected systems have masses comparable to typical star-forming galaxies.} All UBV-selected galaxies at $z=0.5$ are shown as dots, colored by their logarithmic dust mass, with their stellar mass on the $x$-axis and their molecular gas mass on the $y$-axis.  For context, we plot as 2D KDE distributions the associated values for non UBV-selected systems, divided by whether they are star forming (\eq\ref{eq:ssfrt1}; blue), transitioning (\eq\ref{eq:ssfrt2}; yellow), or quenched (\eq\ref{eq:ssfrt3}; red). }
    \label{fig:UBV}
\end{figure}

Applying this selection criterion to our sample of galaxies at $z=0.5$ produces the population colored red in the top plot of \fig \ref{fig:UBV}.  The left-hand panel shows all of our galaxies in UBV-space as individual points, while the red colored dots are those galaxies that satisfy the `PSB-like' UBV-selection criteria (\eq\ref{eq:UBVSelect}).  The right-hand plot of \fig \ref{fig:UBV} shows the same sample using alternative definitions; namely the 4000\r{A} break (D$_n$4000) and Lick index (H$\delta_A$).  The 4000\r{A} break, commonly called the D$_n$4000 index, is used largely to determine a galaxies' current state of star formation \citep{dn4000}, and is defined as the ratio between the average flux density (in erg$~$s$^{-1}$cm$^{-2}$Hz$^{-1}$) between 4000-4100\r{A} and  3850-3950\r{A}. 

In contrast, H$\delta_A$ is a Lick index developed as a standardized measure of the equivalent width of the Balmer H$\delta$ absorption line \citep{hdelta}, which is strongest for galaxies with stellar populations dominated by A-type stars, and is usually interpreted as evidence of a burst of star formation that ended 0.5-1.5 Gyr prior to observation. To calculate this, a pseudo-continuum is defined by calculating the average flux density bluewards of the line (4017\r{A}-4057\r{A}) and redwards of the line (4153\r{A}-4193\r{A}), fitting a straight line between both, then using that to integrate over the absorption line (defined between wavelengths 4083.5\r{A} and 4122.25\r{A}).  The subscript ``A" is used to define a wider bandpass, around 40\r{A}, which is more suited for measuring absorption in A-star dominated systems with intrinsically broader absorption \citep{hdelta}.

In the bottom panel of \fig\ref{fig:UBV}, we plot all UBV-selected galaxies at $z=0.5$, colored by their logarithmic dust mass, with their stellar mass on the $x$-axis, and their molecular $(\hh)$ gas mass on the $y$-axis.  For context, we plot as 2D KDE distributions the associated values for non UBV-selected systems, colored by whether they are star forming (\eq\ref{eq:ssfrt1}; blue), transitioning (\eq\ref{eq:ssfrt2}; yellow), or quenched (\eq\ref{eq:ssfrt3}; red).  From this plot, we can see that while the majority of UBV-selected galaxies have masses comparable to typical star-forming galaxies, the most dust-rich systems instead have higher molecular masses, comparable to typical quenched galaxies.

\section{Results}
\label{sec:results}
To investigate the mechanisms that could give rise to observations of massive quenched galaxies harboring significant molecular gas reservoirs, we divide our theories into three hypotheses, originally discussed in the introduction (\S \ref{sec:intro}), to individually study their relative contributions across cosmic time.  We now discuss our results in the context of these hypotheses in turn.

\subsection{Hypothesis A: Temporary Quiescence}
\label{ssec:HA}
Our first line of investigation is to study whether or not massive galaxies that are quenching (or quenched), with significant reservoirs of molecular gas, are simply in a dormant state and not permanently quiescent; i.e., are these galaxies destined for rejuvenation \citep{remus2023, Kimmig25}?  Indeed, as discussed in the introduction (\S\ref{sec:intro}), many theoretical and observational studies have concluded that rejuvenation can play an important role in galaxy evolution \citep{Lotz21, remus2023, Kimmig25, harrold2025, colibre, cenci25, setton2025}.

\begin{figure}
    \centering
    \includegraphics[width=\columnwidth]{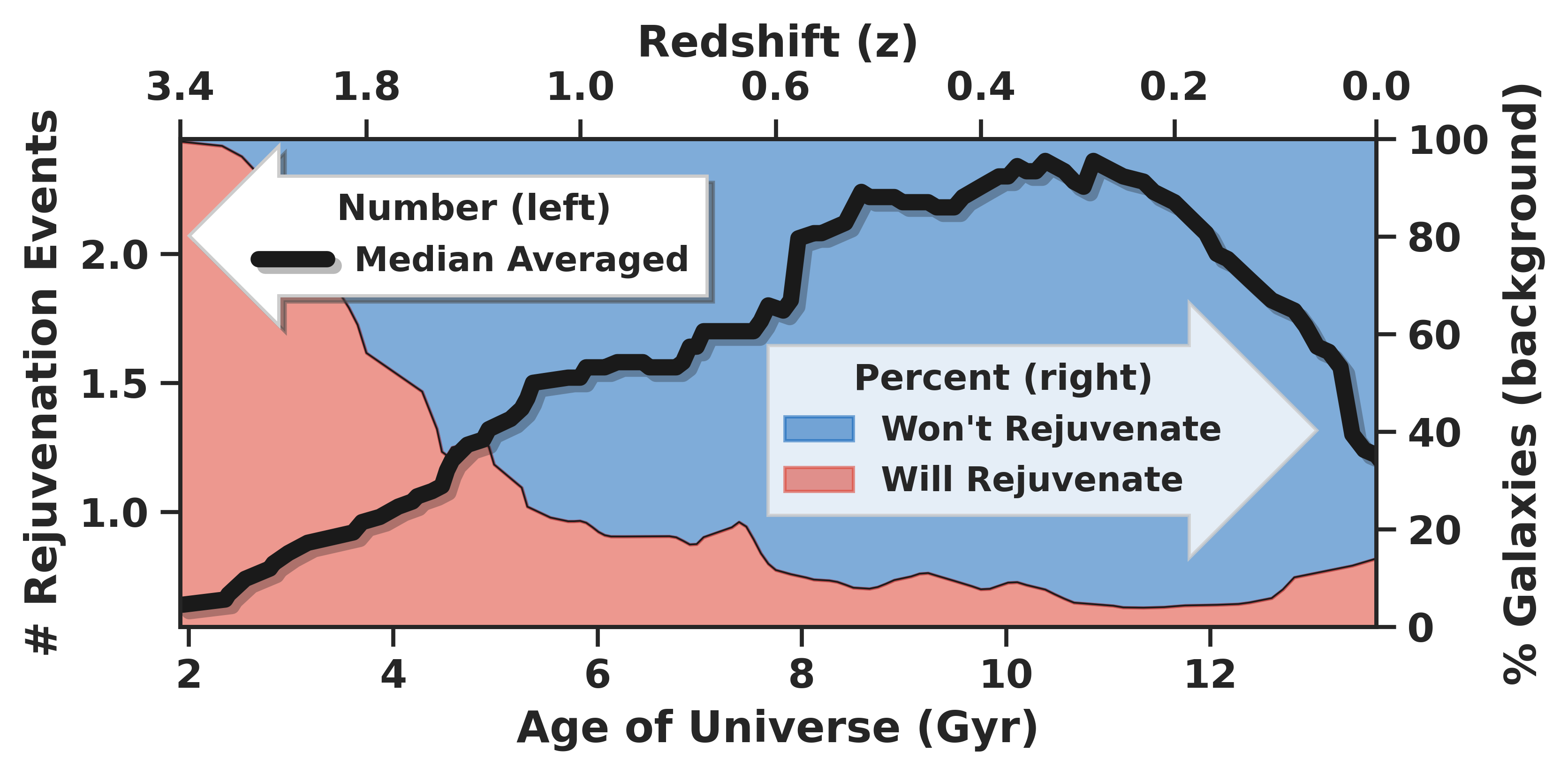}
       \caption{{\bf While the median number of rejuvenation events tends to increase toward present day, the  fraction of $\bm{M_*>3\times 10^9M_\odot}$ galaxies with $\bm{\ssfr\times\tH<0.7}$ that will rejuvenate instead decreases over the same redshift range.} The number of rejuvenation events at each snapshot is shown as a black line (left $y$-axis).  Here, we see low values at high-$z$, due to the rarity of high-mass transitioning and quenched systems at these epochs, as described in \S\ref{ssec:selectGal}/\ref{ssec:select}.  Additionally, we see another turnover at low-$z$, occurring around $z\sim 0.4$ ($\tH\sim 10$Gyr); this arises because galaxies identified within the last few Gyr of cosmic time simply have not had enough time to undergo the $\sim2$Gyr rejuvenation process.  
       In contrast, the fraction of our massive transitioning or quenched galaxies that will rejuvenate, indicated by the red region (right $y$-axis) is highest at high-$z$, and decreases monotonically towards $z=0$.  The reason for this trend is that rejuvenation is primarily driven by galaxy interactions, which in this redshift/mass range are most common around $z\sim 3$, with decreasing rates towards lower $z$.}
    \label{fig:rejuv}
\end{figure}

To define a rejuvenation event, we loosely follow the formalism of \citet{mergers}: here, a galaxy is defined to have rejuvenated if ``after being quenched for a sufficient time, it has its sSFR return back over the star-forming cut".  Specifically, \citet{mergers} defines a rejuvenating galaxy as one that after its $\ssfr$ falls below their quenching threshold of $0.2/\tH$, it remains below $\tH^{-1}$ for at least $0.2\tH$ before having its sSFR rise back above $\tH^{-1}$.  Applying this to our sample, we instead define a galaxy to be a candidate for rejuvenation only if it has $M_*>3\times 10^9M_\odot$, and is either transitioning or quenched (\eq\ref{eq:ssfrt2}/\ref{eq:ssfrt3}) for more than one snapshot.  

We then say that such a galaxy will have rejuvenated if it eventually rises back above the star-forming threshold (\eq\ref{eq:ssfrt1}) for multiple snapshots, and we say it has quenched if it falls below our quenched threshold for more than five consecutive snapshots, or if if it reaches $z=0$ without rejuventing.  While the exact time between snapshots depends on redshift, they are evenly spread out with respect to the logarithmic scale factor $(\log{a})$, where the scale factor is calculated as a function of redshift as $a\equiv 1/(1+z)$.  In contrast to \cite{mergers}, we soften the requirement needed for the galaxy to become eligible for rejuvenation, as our goal here is not to focus exclusively on strongly quenched galaxies, but rather to broadly quantify the likelihood of rejuvenation in any low star-forming and high-mass galaxy.

\begin{figure}
    \centering
    \includegraphics[width=\columnwidth]{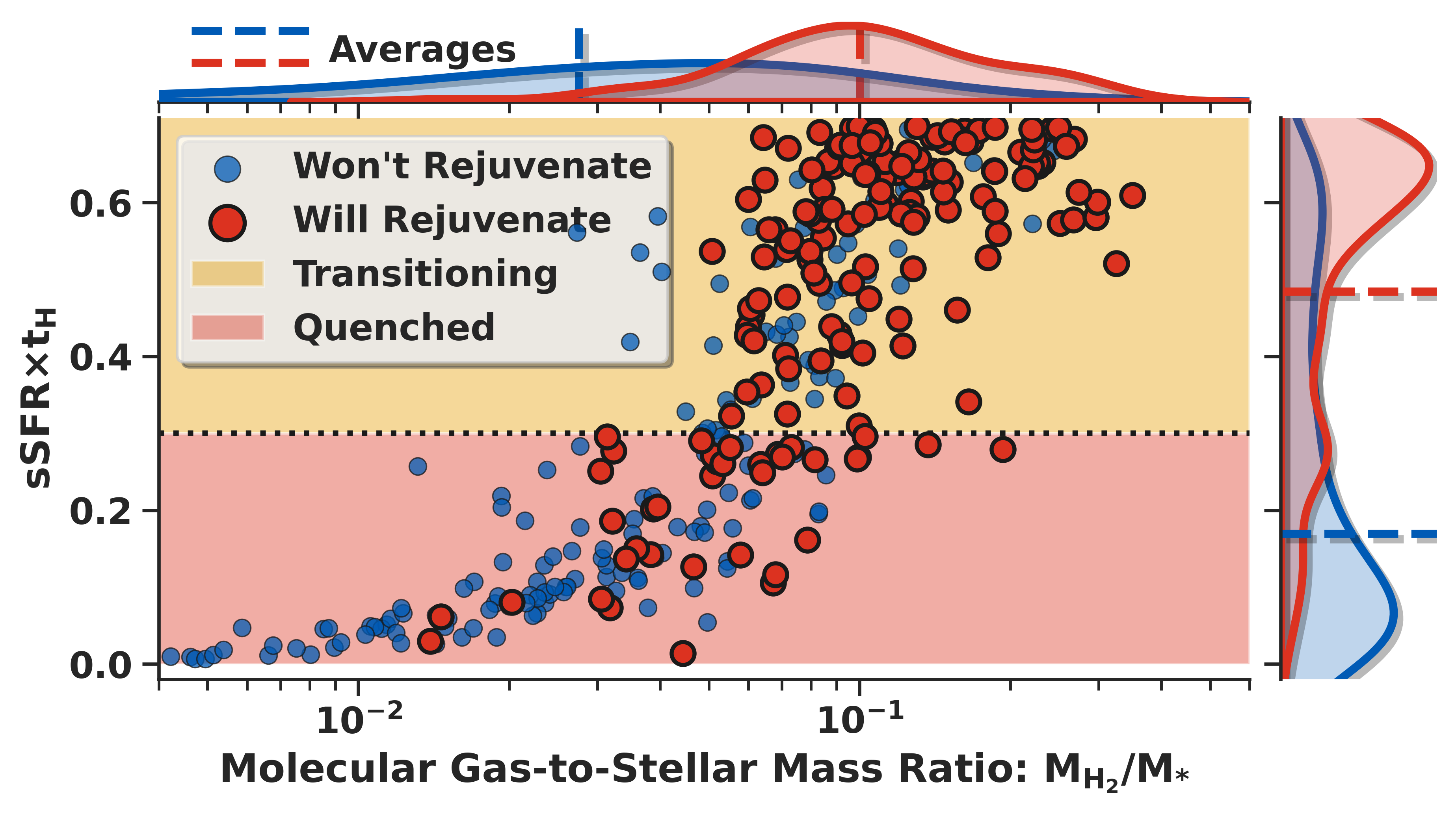}
    \caption{{\textbf{Rejuvenating galaxies are significantly more star-forming and gas-rich than their non-rejuvenating counterparts for $\bm{z=0-3.5}$}}. Every point represents a $M_*>3\times 10^9M_\odot$ galaxy, either transitioning or quenched (\eq\ref{eq:ssfrt2}/\ref{eq:ssfrt3}), colored red if it will rejuvenate and blue if not, with its molecular gas-to-stellar mass ratio shown on the $x$-axis, and its associated value of $\ssfr\times\tH$ on the $y$-axis.  On the top/right sides of the plot we also show the marginal distributions for each population as a KDE plot, with the stellar mass weighted averages shown as dotted lines.  We see from this that the majority of rejuvenating galaxies have $\mhh/M_*>0.1$ (solid black line), implying they are gas-rich, and are predominantly `transitioning' (yellow region; \eq\ref{eq:ssfrt2}), while those that will not rejuvenate are primarily quenched (red region; \eq\ref{eq:ssfrt3}) }
    \label{fig:rejuvComb}
\end{figure}
\begin{figure*}
    \centering
    \includegraphics[width=\linewidth]{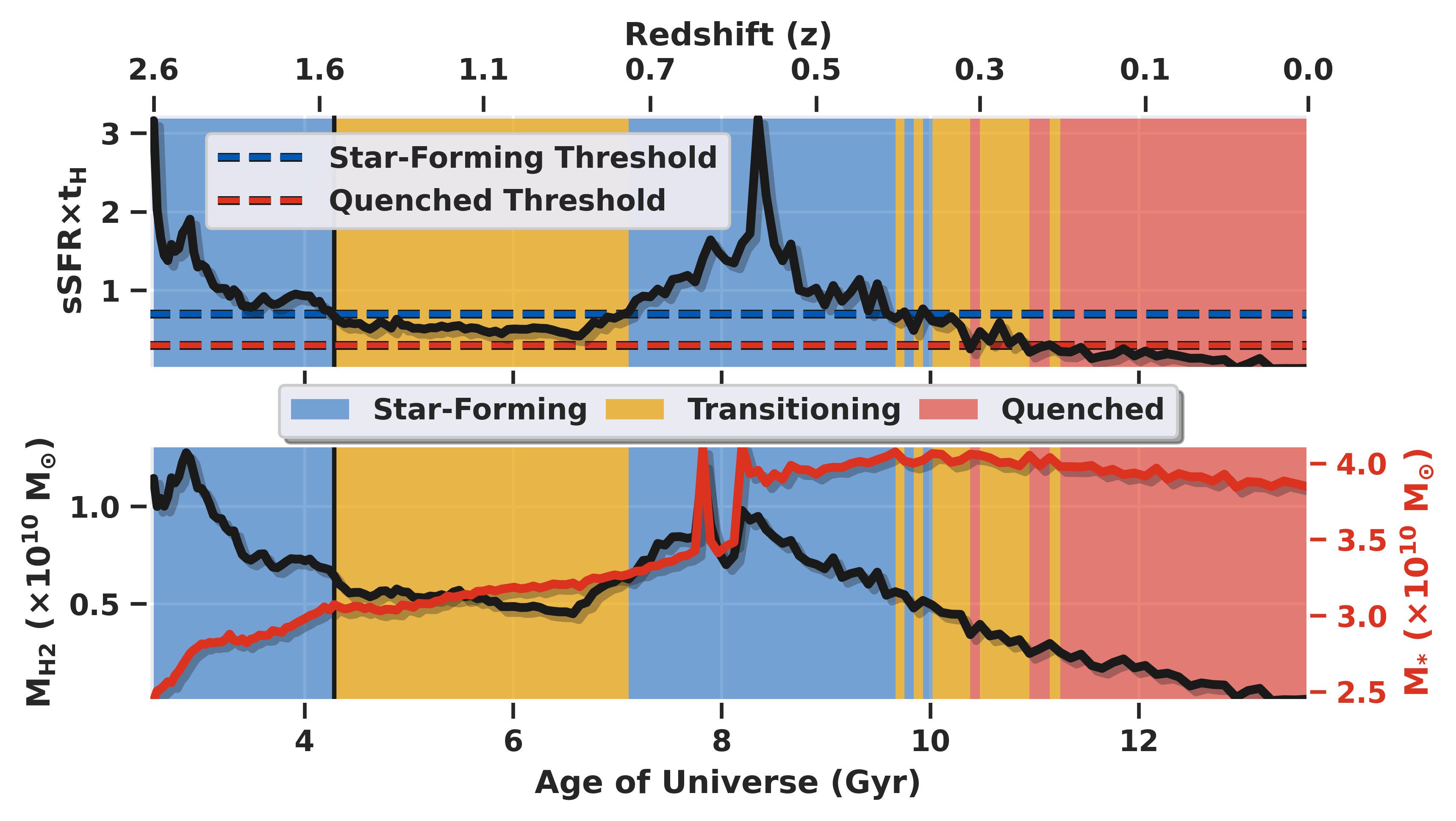}
    \caption{{\bf Example of a rejuvenating galaxy}, selected at $z\approx 1.5$ (vertical black line; $\tH\sim 4.3$\,Gyr), based on it satisfying both our conditions; (a) massive $M_*>3\times 10^9M_\odot$ and (b) $\ssfr\times \tH<0.7$ (transitioning or quenched), with the background color set to the star-forming classification of the galaxy based on \eq \ref{eq:ssfrt1}-\ref{eq:ssfrt3}.  While the top panel shows the evolution of $\ssfr\times \tH$, the bottom panel shows the corresponding evolution of molecular gas mass (left/black) and stellar mass (right/red), both of which show a gradual evolution at high redshift, followed by a dramatic increase at around $z\approx 0.6$ driven by a major merging event.  The drop-off after this and subsequent rise show that the merging galaxy initially passes through our main galaxy, but, now bound within its potential well, later falls back and ultimately merges with it.  This scenario was confirmed in \app \ref{app:merge} via density projection plots of the galaxy at different stages in its evolution.}
    \label{fig:rejuvCase}
\end{figure*}

The results of this analysis are shown in \fig \ref{fig:rejuv}, where for each snapshot we show the median number of rejuvenation events (left), and the percent of galaxies that will (red) and won't (blue) rejuvenate (right).  The left-hand $y$-axis shows as a black line the median averaged number of rejuvenation events that occur at each snapshot, where at each snapshot we take a median of the values within a moving window centered on that snapshot in order to smooth the data.  In contrast, the right-hand $y$-axis shows the percent of our tracked galaxies at each snapshot that eventually will (red region) and won't (blue region) rejuvenate when tracked forward in time.  Similar to the black line, these regions have been smoothed using a 1D Gaussian ($\sigma=2$) filter from {\sc scipy.ndimage}\footnote{\url{https://docs.scipy.org/doc/scipy/reference/generated/scipy.ndimage.gaussian_filter1d.html}}, which replaces points with a weighted average of nearby points via a Gaussian kernel.

Examining \fig \ref{fig:rejuv}, we see that while rejuvenation can be an important effect at high redshift, it reduces in relative likelihood at low redshift (as also found in \citealt{Szpila25}).  To expand on this, we see that while at high redshift the occurrence of rejuvenations is the highest, the sample size is also the smallest due to the lack of massive galaxies at high-$z$ (as discussed in \ref{ssec:massLimit}).  In contrast, as redshift decreases, the number of $M_*>3\times 10^9M_\odot$  transitioning/quenched (\eq \ref{eq:ssfrt2}/\ref{eq:ssfrt3}) galaxies that will rejuvenate increases towards a maximum value around $z\sim 0.4$, then again falls towards the present day ($z=0$).  The reason for this trend is two-fold.  First, as will be shown later in this section, rejuvenation is primarily driven by galaxy interactions like mergers, which are most common for massive galaxies at $z\sim 3$ \citep{Conselice07}.  In addition to the rarity of galaxy interactions capable of reigniting star formation in the local universe, the drop off at $z<0.4$ could also be explained by the $\sim$Gyr timescales needed to reignite star-formation, so this turnover can also be thought of as an artifact arising from the lack of time needed to rejuvenate in the early universe.  In other words, while such galaxies may still rejuvenate in the future, not enough time has passed for them to have rejuvenated by the present day.

Given the results of \fig \ref{fig:rejuv}, it becomes instructive to ask what physical differences exist between those galaxies in our sample that rejuvenated or not.  This is accomplished in \fig \ref{fig:rejuvComb}, in which we include all galaxies from \fig \ref{fig:rejuv} on a plot of molecular gas-to-stellar mass fraction vs $\ssfr\times\tH$, for all snapshots from $z=0-3.5$, colored only by whether they will (red) or won't (blue) rejuvenate.  On the top/right sides of \fig \ref{fig:rejuvComb} we show the marginal distributions for each population as a KDE plot\footnote{A kernel density estimate (KDE) plot is a method for visualizing the distributions in a dataset, analogous to a histogram, with the important difference that instead of discrete bins observations are smoothed with a Gaussian kernel, producing a continuous density estimate such that the area under the curve is unity.}, calculated using the python-based {\sc matplotlib} extension called {\sc seaborn} \citep{matplotlib, seaborn}, with the stellar-mass weighted averages shown as dotted lines.

Examining the distributions in \fig \ref{fig:rejuvComb} along the $x$-axis, we see immediately that the galaxies that will rejuvenate are much more molecular gas-rich relative to those that won't, with $\sim50\%$ of rejuvenating galaxies having $\mhh/M_*>0.1$ (12\% of non-rejuvenating systems satisfy this), while $\sim90\%$ of rejuvenating galaxies have $\mhh/M_*>0.05$ (in contrast to 40\% of non-rejuvenating systems), consistent with observational studies of gas-rich quenched galaxies \citep[e.g.][]{french_2015, rowlands}.

Examining the marginal distributions from \fig \ref{fig:rejuvComb} along the $y$-axis shows that rejuvenating galaxies are significantly more star forming than their non-rejuvenating counterparts for $z=0-3.5$.  The background color of the plot is shaded to represent their star-formation classification according to \S\ref{ssec:select}, where the yellow background corresponds to `transitioning' (\eq\ref{eq:ssfrt2}), while the red background is `quenched' (\eq\ref{eq:ssfrt3}).  We see that galaxies that won't rejuvenate have values of $\ssfr\times\tH$ skewed towards the smallest values, with an average value that would fall in our `quenched' category (\eq\ref{eq:ssfrt3}), implying that these systems are fully quiescent, while those that will rejuvenate have the opposite trend, with an average value in our `transitioning' regime.  Specifically, we find that while $\sim80\%$ of rejuvenating systems will reside in the transitioning regime (\eq\ref{eq:ssfrt2}), only around 35\% of non-rejuvenating systems will satisfy this (65\% fully quenched).  Taken as a whole, \fig\ref{fig:rejuvComb} shows that the galaxies that will rejuvenate are both (a) rich in molecular gas (relative to stellar mass) and (b) are not yet fully quenched.

To probe deeper, we show the individual star formation history (SFH) for an example rejuvenating galaxy in \fig \ref{fig:rejuvCase}.  This galaxy was selected at $z\approx 1.5$ (vertical black line; $\tH\sim 4.3$\,Gyr), based on it satisfying both our conditions for having $M_*>3\times 10^9M_\odot$ and being transitioning (\eq \ref{eq:ssfrt2}), with the background color set to the star-forming classification of the galaxy based on \eq \ref{eq:ssfrt1}-\ref{eq:ssfrt3}.  The galaxy was tracked forward in time using descendant tracking, and backward to $\tH\sim 2.6$\, Gyr using progenitor tracking (see \S\ref{ssec:prog} for details).  The bottom panel shows the evolution of H$_2$ mass (left/black) and stellar mass (right/red), showing that this galaxy has $M_*>3\times 10^9M_\odot$ for all points in time.  

The top panel shows the corresponding evolution of $\ssfr\times\tH$ as a solid black line, showing how after initially being identified as transitioning at $\tH=4.3$Gyr, we say the galaxy has rejuvenated approximately 2.8Gyr later at $\tH=7.1$Gyr.  This rejuvenation event is followed by a rapid increase in SFR, followed by a complete quenching event, which occurs in our formalism at $\tH=11.2$ Gyr.  Furthermore, this rejuvenation event in its star formation is seen to follow a dramatic increase in both the galaxy's molecular gas content and stellar mass at around 8 Gyr (bottom row of \fig \ref{fig:rejuvCase}), indicating a major gas-rich merging event occurring here. Confirmation of this scenario was achieved by examining the projection plots of our target galaxy over time, shown in \app \ref{app:merge}, in which we can visually observe the companion galaxy passing through our target system and eventually merging with it.  More details about this interaction and the associated projection plots are shown in \app \ref{app:merge}.

In summary, by examining the example rejuvenating galaxy in \fig \ref{fig:rejuvCase}, we learned that rejuvenation events can be triggered by a gas-rich major merging event.  In \fig \ref{fig:rejuvComb}, we combined our sample of high-mass and low star-forming galaxies over all $z=0-3.5$ to show that those galaxies that will rejuvenate are more gas rich and have significantly higher levels of star formation than those that do not rejuvenate.  With \fig \ref{fig:rejuv} we analyzed the redshift dependence associated with rejuvenation, and showed that the highest probability of rejuvenation occurs at high redshift.  Despite having the lowest sample size of galaxies that fit our selection criteria at high-$z$ due to the effects of our high-mass cut, as discussed in \S\ref{ssec:massLimit}, the abundance of major mergers at high-$z$ is most likely what drives the high probability of rejuvenation here.  In contrast, while the largest sample of massive galaxies are found at low redshift, the probability of rejuvenation here is also the smallest. While such galaxies may still rejuvenate in the future, the salient point is that they have not yet had time to rejuvenate by $z=0$.

Putting it all together paints a picture in which among the transitioning or quenched massive galaxies, those that will rejuvenate are those that are relatively more molecular ($\hh$) gas rich and have not yet fully quenched.  While these findings could potentially explain why galaxies with persistent molecular gas reservoirs could appear to have anomalously low SFRs, such as those from \cite{suessSquiggle} ($0.5<z\leq 0.9$), it does not resolve the question of why their molecular gas seems to disappear on extraordinarily fast ($\sim150$ Myr) timescales \citep{setton2025}.

\subsection{Hypothesis B: Incorrect Alpha-CO}
\label{ssec:HD}
\begin{figure*}[hbt!]
    \centering
    \includegraphics[width=\linewidth]{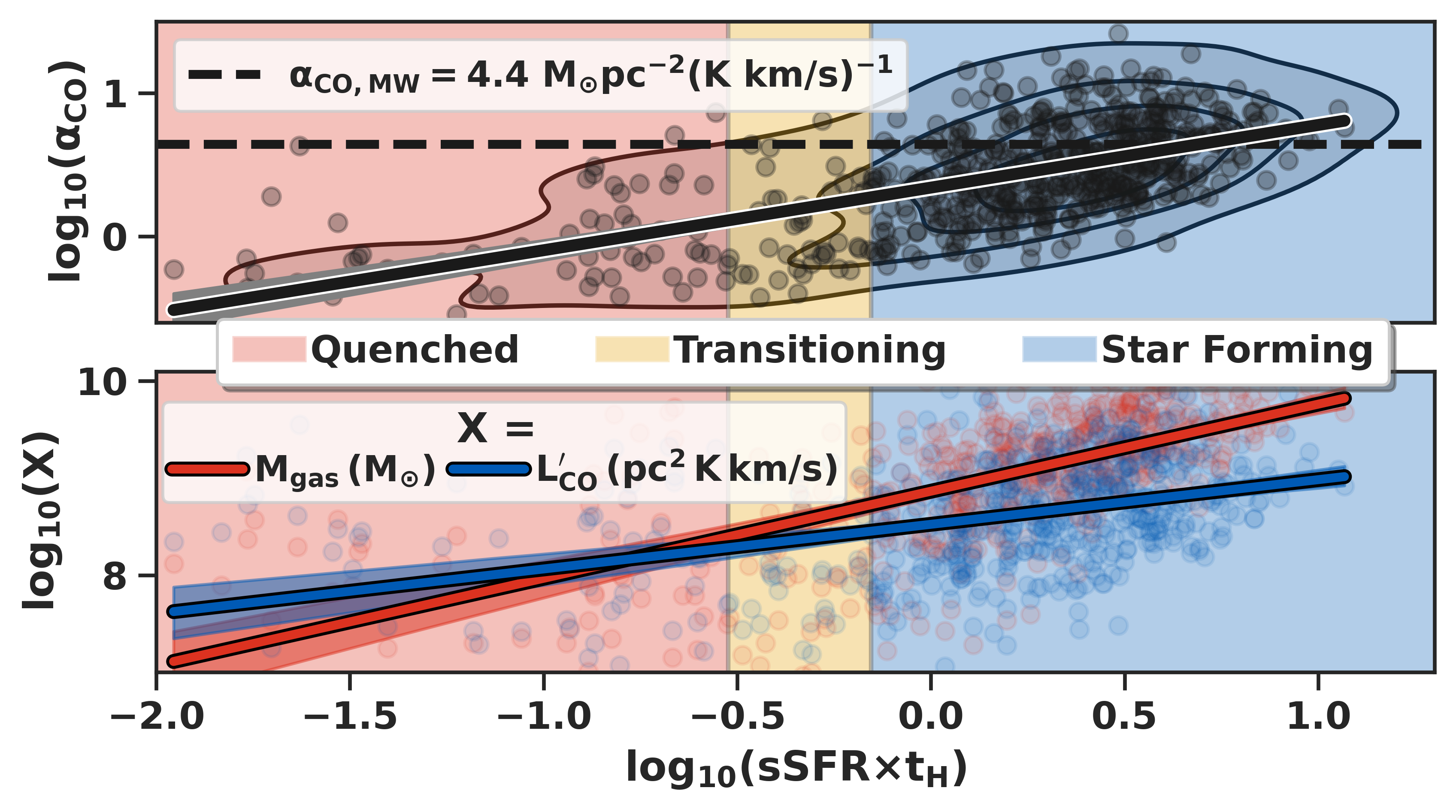}
    \caption{{\bf (\underline{Top}) The value of $\bm{\aco}$ increases with $\bm{\ssfr\times \tH}$ such that systems with low levels of star-formation will typically have a less than MW-like value of the CO-$\bm{\hh}$ conversion factor ($\bm{\aco<\acomw}$), implying observers who assume $\bm{\aco=\acomw}$ would overestimate $\bm{\mhh}$.}  All massive $(M_*>3\times 10^9M_\odot)$ galaxies are shown here for $z=0,1$, with their logarithmic value of $\ssfr\times\tH$ on the $x$-axis and $\aco$ on the $y$-axis (in units $\mathrm{M_{\odot}\, pc^{-2}(K~km/s)^{-1}}$).  As in previous plots, the background shading corresponds to our star-forming classification scheme from \S\ref{ssec:select}, in which the blue background corresponds to galaxies that are star-forming (\eq\ref{eq:ssfrt1}), yellow for transitioning (\eq\ref{eq:ssfrt2}), and red for quenched (\eq\ref{eq:ssfrt3}).  The black line shown here is a linear regression model fit to our data, while the shaded region surrounding it is the associated 95\% confidence interval, estimated using bootstrap resampling.  From this, we see a trend in which $\aco$ is lower than $\acomw$ for quenched and transitioning galaxies, and only surpasses this for the most actively star-forming galaxies ($\ssfr\times\tH>5$). {\bf (\underline{Bottom}) The reason that higher levels of star-formation correlate with higher $\bm{\aco}$ stems from the fact that as a function of increasing $\bm{\ssfr\times \tH}$, the gas mass (red line) in a galaxy will increase at a faster rate than the integrated CO luminosity (blue line).}  While higher levels of star-formation are correlated with both an increase in $\hh$ and CO abundances, the decreasing temperatures that result from molecular gas shielding in dense star-forming regions lowers the rate at which the CO luminosity ($\lco$) will increase with respect to the gas mass ($\mgas$).}
    \label{fig:alphaCO}
\end{figure*}
Our second hypothesis explores whether an incorrectly assumed value of the CO-to-H$_2$ conversion factor ($\alpha_\text{CO}$) could be the root cause of the seemingly abundant molecular gas reservoirs observed in massive quenched galaxies.  Since $\hh$ resides primarily in cool giant molecular clouds \citep[GMCs;][]{Mckee07}, yet only emits directly at high temperatures, it is nearly invisible via direct emission in these environments and thus must be traced using the next most abundant molecule, carbon monoxide (CO).  After measuring the velocity-integrated CO luminosity in a GMC $(L_{\text{CO}}')$, the corresponding molecular gas mass $(\mgas)$ can be estimated by using the CO-to-H$_2$ conversion factor ($\aco$), defined as $\aco=\mgas/L_{\text{CO}}'$ \citep{Bolatto_2013, alphaCO}.  Owing to the large number of studies of this factor in the Milky Way that converge on a characteristic value of $\acomw=4.4\ \text{M}_\odot \text{pc}^{-2} (\text{K}~\text{km/s})^{-1}$ \citep{Narayanan2013,Bolatto_2013}, extragalactic studies often adopt this Milky Way value for $\aco$ across all galaxies.

Since systematic variations of $\aco$ with various ISM properties are both expected and observed \citep{Strong04,narayanan12, Sandstrom13}, in this section we will explore how the common assumption that $\aco=\acomw$ will impact the inferred molecular gas in massive gas-rich quiescent galaxies.  All values needed to calculate $\aco=\mgas/L_{\text{CO}}'$ are computed in post-processing using the \slick code \citep{slick}, which derives model CO luminosities by splitting up each gas particle into concentric radial zones \citep{popping} and computing thermal, radiative, and statistical equilibrium throughout the sub-grid model using \despotic\!\!\footnote{\despotic is a code to Derive the Energetics and SPectra of Optically Thick interstellar Clouds and can be found at \url{https://bitbucket.org/krumholz/despotic/src/master/}.}\citep{despotic}.  In this formalism, each gas particle can be thought of as representing a giant molecular cloud (GMC).  

The molecular gas $(\mgas)$ in our model assumes a GMC composition of molecular Hydrogen (H$_2$) and atomic Helium (He), in which the Helium mass is 40\% of H$_2$, so that $\mgas \equiv \mhh+M_{\text{H}_e}=\mhh(1+0.4)=7\mhh/5$.  Then to calculate $\aco$ for a galaxy in our simulation, we computed the total molecular gas mass of each gas particle that converged in post-processing, add their velocity-integrated CO luminosities, and obtain their ratio to get $\aco$.  

To investigate the relationship between the resulting $\aco$ of a galaxy and its level of star-formation, we show both quantities on a logarithmic scale for all $z=0,1$ massive ($M_*>3\times 10^9M_\odot$) galaxies in \fig \ref{fig:alphaCO}.  Here, the top panel shows $\aco$ on the $y$-axis and $\ssfr\times\tH$ on the $x$-axis, while the bottom panel on the $y$-axis instead shows the numerator/denominator of $\aco$ in red/blue.  As in the last section, the background shading corresponds to our star-forming classification scheme from \eq\ref{ssec:select}, so that the blue background corresponds to galaxies that are star-forming (\eq\ref{eq:ssfrt1}), yellow for transitioning (\eq\ref{eq:ssfrt2}), and red for quenched (\eq\ref{eq:ssfrt3}).  In both panels, we fit a linear regression to the data and show the resulting relationship as a solid line, surrounded by a 95\% confidence interval (estimated using bootstrap resampling).  From the top panel in \fig \ref{fig:alphaCO}, we see that as a galaxies level of star-formation increases (quantified by $\ssfr\times \tH$; c.f. \S\ref{ssec:select}), as does its value of $\aco$.  Furthermore, we see that all quenched, transitioning, and weakly star-forming galaxies will tend to have a value of $\aco$ less than $\acomw$, implying that the assumption that $\aco=\acomw$ in most recently quenched massive galaxies will cause an overestimated value of $\mhh$.  In fact, the linear regression shows that $\aco>\acomw$ only for galaxies with $\ssfr\times \tH>5$.

In the bottom panel of \fig \ref{fig:alphaCO}, we show the linear regressions of both the numerator $(\mgas)$ and denominator $(\lco)$ of $\aco$ as red and blue lines, respectively, in order to better visualize the relationships between $\lco$ and $\mgas$ vs $\ssfr \times \tH$.  The CO luminosity, in optically thick gas, scales with the molecular gas velocity dispersion and temperature \citep{narayanan12,Narayanan2013}.  Galaxies that are quenched in our models tend to have lower column density ISM conditions, which result in warmer gas temperatures.  This results in higher $\lco$ per unit gas mass, driving the conversion factor below the Milky-Way mean value. We can broadly conclude that the recently quenched galaxies will tend to have $\aco<\acomw$, such that any observations that assume $\aco=\acomw$ may overestimate the $\hh$ mass.  

\subsection{Hypothesis C: Dust Obscuration}
\label{ssec:HC}
Our third hypothesis is that massive galaxies that appear to be quenched, but are rich in molecular gas, are actually harboring dust-obscured star formation, and are not actually quenched at all \citep[e.g.,][]{Baron22, Baron2023, pengPei, cenci25}.  Specifically, \cite{Baron2023} analyzed NOEMA CO(1-0) observations of 15 post-starburst galaxies with a multiwavelength approach, finding that while some regions of the galaxy are indeed quenched, other parts of the galaxy are instead forming stars in highly obscured regions.

In contrast to the previous two hypotheses, in this section we instead select our sample of galaxies based on their colors instead of their SFR, as we are testing whether or not star-forming galaxies can mimic signatures of quenched galaxies in their observational spectra (in other words: in order to fully test this hypothesis, we need to select low star-forming galaxies in a manner similar to observers).  To accomplish this, we utilize the UBV-classification scheme described in \S\ref{ssec:ubvSelect}.

\begin{figure*}
    \centering
    \includegraphics[width=\linewidth]{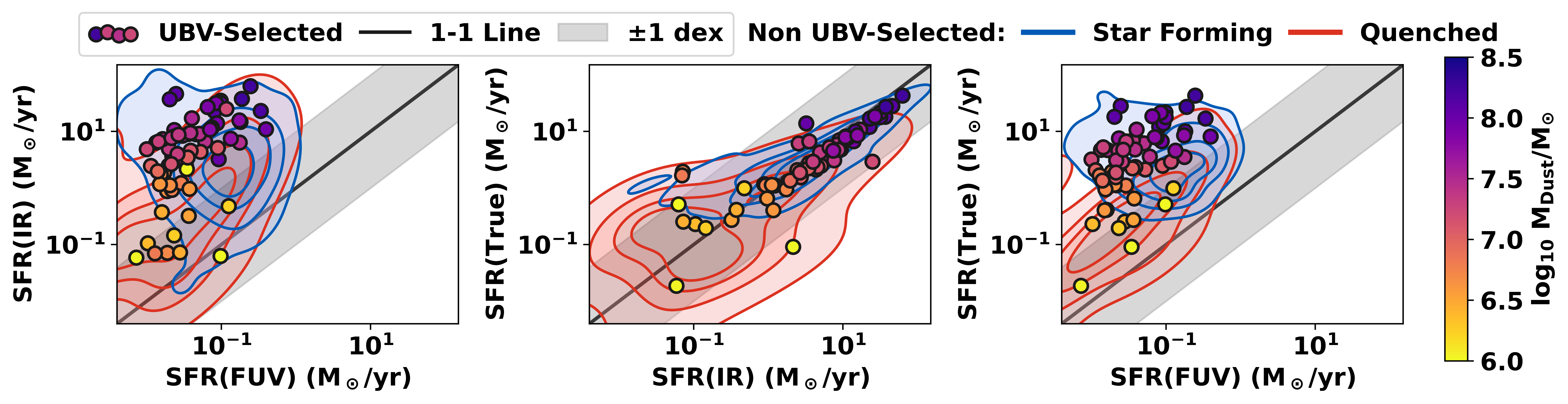}
    \caption{{\bf UBV-selected massive galaxies at $\bm{z=0.5}$ have an IR-based SFR that accurately traces their true SFR, yet a FUV-based SFR that significantly underestimates it.}  The left plot shows that among the UBV-selected galaxies, those that are most dust-rich will have an excess of dust-obscured star formation, which shifts the emitted UV starlight into the IR.  As a result, the IR luminosity traces the SFR in these systems (middle plot), while the UV luminosity severely underestimates it (right plot).  In contrast, the UV-based SFR works best for the most dust-poor systems, for which the IR-based SFR is the least reliable for.}
   \label{fig:IR_FUV}
\end{figure*}

After identifying which galaxies will satisfy our UBV-selection (\eq\ref{eq:UBVSelect}) at $z=0.5$ (red points in \fig \ref{fig:UBV}), we continue to investigate how the SFR derived from the IR deviates from the SFR derived from the FUV, as well as how both compare to the true SFR.  Since most SFR indicators are derived from rest-UV or optical measurements, these methods are likely to yield lower SFR estimates than expected in dust-rich systems, as a significant portion of the high-energy radiation from the galaxy is absorbed by dust and re-emitted in the infrared.  To test this hypothesis, we take our sample of UBV-selected galaxies (red points in \fig \ref{fig:UBV}) and calculate the SFR of these galaxies derived from their mock IR and far-UV photometry, to compare with the true SFR of the galaxies, with corresponding results shown in \fig \ref{fig:IR_FUV}.

Following \cite{murphy}, we define L$_{\text{IR}}$ as the integrated luminosity of our dust-attenuated SED from 8-1000$\mu$m, and convert it to an IR-based star-formation rate using $\sfr_{\text{IR}}/(\text{M}_\odot/\text{yr})=3.88\times 10^{-44}\text{L}_{\text{IR}}/(\text{erg/s})$.  Similarly, we define L$_{\text{FUV}}$ by convolving the SED with a {\sc GALEX\,\,}FUV transmission curve, and compute the corresponding FUV-based star-formation rate as $\sfr_{\text{FUV}}/(\text{M}_\odot/\text{yr})=4.42\times 10^{-44}\text{L}_{\text{FUV}}/\text{(erg/s)}$.

The results of applying this analysis to our sample of UBV-selected galaxies at $z=0.5$ are shown in \fig \ref{fig:IR_FUV}.  Here, we show all UBV-selected selected systems (with $M_*>3\times 10^9M_\odot$), colored by their dust mass.  Additionally, for context, we also show as contours the non UBV-selected systems, split by whether they are star-forming (\eq\ref{eq:ssfrt1}; $\ssfr\times\tH>0.7$) or quenched (\eq\ref{eq:ssfrt3}; $\ssfr\times\tH<0.3$).  In \fig \ref{fig:IR_FUV}, the left-most panel shows a direct comparison of FUV-based SFR vs IR-based SFR, showing an excess of dust-obscured star formation for those UBV-selected systems that are also dust-rich.  The middle plot shows the IR-based SFR vs true SFR, and demonstrates that the dust-rich UBV-selected galaxies have IR luminosities that accurately trace their true SFRs, while the IR luminosity of dust-poor galaxies is a less reliable indicator of true SFR.  Finally, the right-most plot shows FUV-based SFR vs true SFR, and shows that while the UV luminosity severely underestimates the true SFR for all UBV-selected systems, it has the least severe effect for the most dust-poor systems.  In all panels we see that among our UBV-selected galaxies, those that are dust-rich seem to mimic the trends of normal star-forming galaxies, while those dust-poor systems instead tend to have SFRs more similar to quenched systems.

From \fig \ref{fig:IR_FUV}, see that while the IR-based SFR accurately traces the true SFR only in the most dust-rich systems, the FUV-based SFR has the opposite trend, in which the best agreement with true SFR occurs for the most dust-poor systems.  This makes sense given that dust emits primarily in the IR, so the most dust-rich systems in our UBV-selected sample are absorbing all their UV energy, reprocessing it into the higher wavelength IR emission, which then faithfully traces the true SFR.  This is also why the most dust-poor systems have the best agreement between FUV-based SFR and true SFR.  These findings are in contradiction with the assumption that UBV-selection will target relatively dust-free galaxies.  Furthermore, from \fig\ref{fig:UBV}, we can also see that the most dust-rich galaxies also have the largest molecular and stellar masses.

\begin{figure*}
    \centering
    \includegraphics[width=\linewidth]{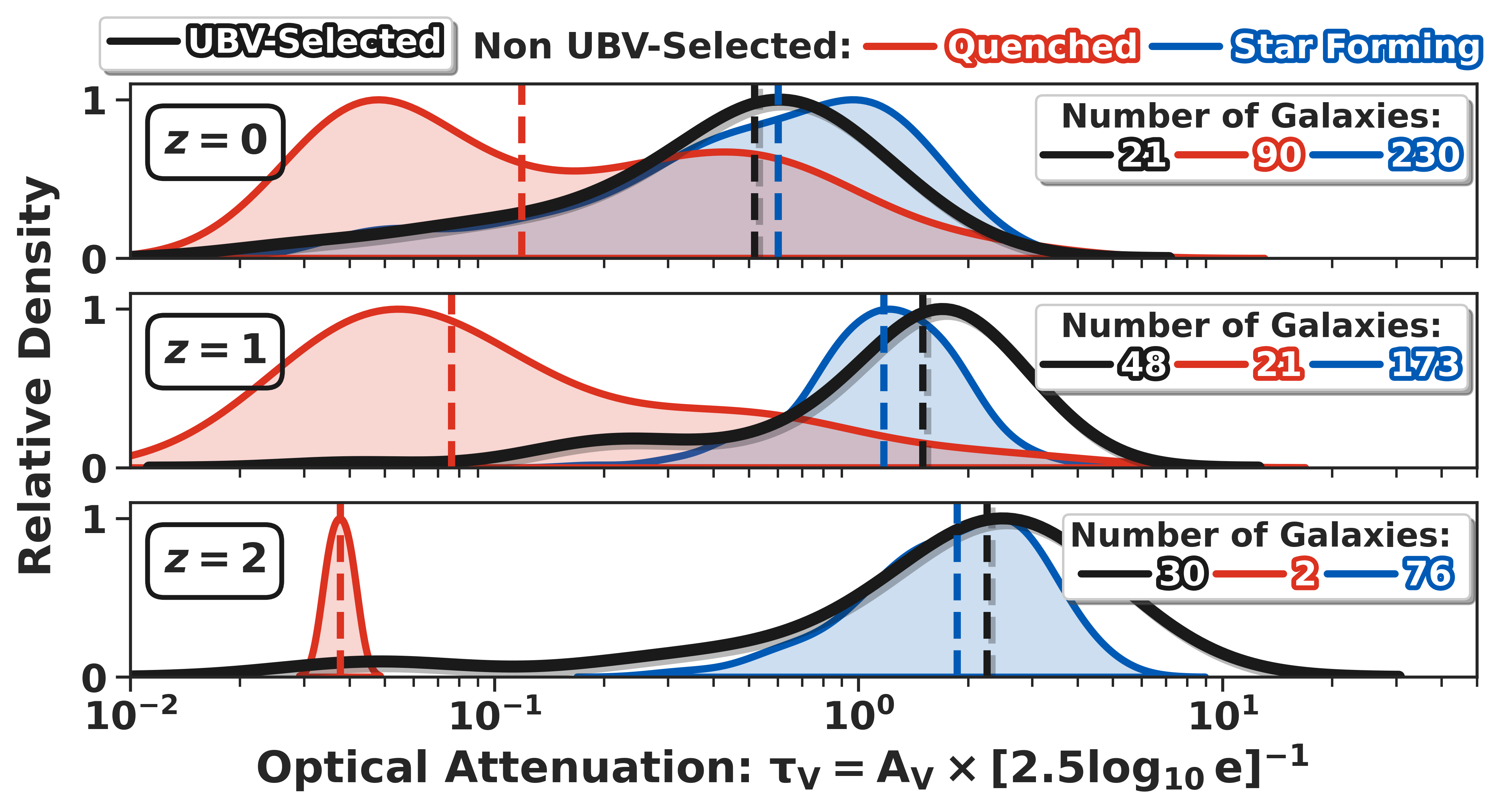}
    \caption{{\bf Massive ($\bm{M_*>3\times 10^9M_\odot}$) galaxies with PSB-like UBV colors (\eq\ref{eq:UBVSelect}) have an optical attenuation similar to typical star-forming galaxies (\eq\ref{eq:ssfrt1}), much higher than typical quenched galaxies (\eq\ref{eq:ssfrt3}), with a value of $\bm{\tau_V}$ that increases with redshift.}  Here, we show peak-normalized distributions of the visual optical depth (see legend for galaxy counts), calculated from attenuation curves, for our population of UBV-selected galaxies (black line), for our non UBV-selected star-forming galaxies (blue line), and for our non UBV-selected quenched galaxies (red line), to see how all three populations' extinction properties evolve with time.  The median value for each population is shown as a dashed line.}
    \label{fig:observer}
\end{figure*}

Thus, from \fig\ref{fig:IR_FUV} we have now shown that among those galaxies that satisfy the PSB-like UBV criteria (\eq\ref{eq:UBVSelect}) at $z=0.5$, those that are most massive tend to be significantly more dust obscured than their low-mass counterparts.  We continue by asking how this trend evolves with redshift by investigating how the extinction properties of the three populations shown in \fig\ref{fig:IR_FUV} evolve with redshift.  This is accomplished in \fig \ref{fig:observer}, in which for $z=0,1,2$ we plot the KDE distribution of the visual optical depth $(\tau_V\equiv A_V*[2.5\log_{10}\!{e}]^{-1})$.  We approximated $\tau_V$ by using the value of the total extinction at $\lambda=0.55\mu\text{m}=5500$\r{A}.  As a result, it follows that $\tau_V\equiv \tau_\lambda(0.55\mu\text{m})=\big(\frac{\ln{10}}{2.5}\big)\times A_\lambda(0.55\mu\text{m})$.  Furthermore, each distribution is scaled so that its peak equals one; therefore curve heights do not reflect relative numbers of galaxies, which are instead given in each panel's legend.

Within each plot in \fig \ref{fig:observer}, we show the peak-normalized distributions of the optical attenuation for all UBV-selected galaxies (\eq\ref{eq:UBVSelect}) in our simulation as a black line, those that are not UBV-selected and defined as star-forming according to \eq\ref{eq:ssfrt1} are shown as a blue line, and those that are not UBV-selected and defined as quenched according to \eq\ref{eq:ssfrt3} are shown as the red line.  In \fig \ref{fig:observer}, we see that at all redshifts, the UBV-selected galaxies have an optical attenuation that tracks that of non UBV-selected star-forming systems, and has significantly higher values of $\tau_V$ then the non UBV-selected quenched systems.  Furthermore, we find that the optical attenuation of UBV-selected systems increases with increasing redshift, with a median $\tau_V$ that increases from $0.52$ to $2.25$ from $z=0\to2$, showing how such systems will become more dust obscured at earlier times.  In contrast, the median $\tau_V$ of the non UBV-selected and quenched systems decreases with increasing redshift, dropping from $\tau_V=0.12\to 0.04$ from $z=0\to2$.

Collectively, the results of this section indicate that dust-obscured star formation may reasonably explain why observations are finding molecular gas in seemingly quenched galaxies.  Using observationally-based selection criteria, we analyzed a sample of PSB-like galaxies and have shown that while a SFR derived from the FUV will underestimate the true SFR for all such galaxies, the IR-based SFR instead will accurately trace the true SFR for dust-rich galaxies (yet still underestimate it for dust-poor galaxies).  We also found that such dust-rich galaxies will also have an abundance of gas and stars (\fig\ref{fig:UBV}).  

In summary, this section suggests that the seemingly quenched massive and molecular gas-rich galaxies that exhibit PSB signatures instead contain dust-obscured star formation hidden in the galaxies center; surrounded by a quenched, dusty, and optically-thin outskirt that enshrouds the dust and exhibits PSB signatures.

\newcommand{\firebox}[0]{{\sc FIREbox} }
\newcommand{\fire}[0]{{\sc fire}}
\newcommand{\cenci}[0]{\cite{cenci25} }
\section{Discussion: Comparison to Other Theoretical Models}
\label{sec:discussion}
Large-volume hydrodynamic simulations have had great success in the past in reproducing many of the statistical trends seen in observations \citep{Somerville2015, tng1b, vogelsberger2019, Crain_2023}.  Cosmological simulations such as these thus provide a potentially powerful avenue towards learning how massive galaxies can preserve their molecular gas reservoirs after star formation shuts off.  Throughout this paper we used the \simba simulation to investigate how molecular gas reservoirs could persevere in recently quenched massive galaxies. 

To place our results in the context of previous work, this section reviews other theoretical studies of analogous systems, namely post-starburst galaxies (PSBs) as introduced in the introduction (\S\ref{sec:intro}).  In this section we will focus on two major theoretical studies investigating the origin of molecular gas in PSBs.  First, \cite{psbEagle} used the \eagle simulations to investigate the molecular gas properties of PSBs, and secondly, \citet{cenci25} studied PSBs with the \firebox simulation \citep{firebox}, which is part of the \textit{Feedback in Realistic Environments} (\fire) project \citep{fire, fire2}.

In terms of galaxy selection criteria, \cite{psbEagle} identified their sample of PSB galaxies in \eagle by focusing on massive ($M_*\gtrsim 3.2\times 10^9M_\odot$) galaxies, initially star forming ($\ssfr>5\times 10^{-11}yr^{-1}$), that experienced at least a factor of 5 drop in sSFR within 600 Myr.  In contrast, \cenci instead first used the same UBV-selection criteria applied in \S\ref{ssec:HC} to the central galaxies in \firebox with $M_*\ge 3\times 10^9 M_\odot$ to define their sample of PSB galaxies.  A `true PSB' was then defined to be any such system with $\text{sSFR}<3\times 10^{-11} \textrm{yr}^{-1}$, that also experienced a least a factor of four drop in their SFR since their most recent starburst, while an `impostor PSB' would be any photometrically selected PSB that does not satisfy these additional criteria (i.e. UBV-selected but star-forming systems).  This approach effectively combines aspects of our sSFR-based galaxy selection criteria described in \S\ref{ssec:select} with the photometric selection method from \S\ref{ssec:ubvSelect}.

Using these selection criteria, \cenci found that among their star-forming galaxies, around 2.2\% at $z=0.7,1.0$ would be classified as `imposter PSBs', thus instead represent star-forming systems misclassified as PSBs due to their photometry, and will contain substantial molecular gas reservoirs with depletion times of 600 Myr -- 1.2 Gyr.  In contrast, those identified as `true PSBs' (quenched and UBV selected) will be gas-poor, and will exhibit the longest depletion times.  This is in line with our findings, in which among our UBV-selected systems (\eq\ref{eq:UBVSelect}) at $z=0.5$, those that are also star-forming (\eq\ref{eq:ssfrt1}) will have median gas depletion timescales of 798 Myr, while those that are also quenched (\eq\ref{eq:ssfrt3}) will have significantly longer depletion timescales, with a median of $\sim2$ Gyr.  Similarly, \cite{psbEagle} found that the average PSB in \eagle\! loses $\sim 90\%$ of its star-forming ISM within $\sim$600 Myr following the starburst.

\cenci showed that typical PSBs, despite representing only 3.1\% of $M_*>3\times 10^9M_\odot$ galaxies at $z=0.5, 1$, tend to have molecular gas contents similar to star-forming galaxies of similar mass.  Of these, only $\sim$32\% are classified as `true PSBs'.  Such systems are found to be significantly gas-poor, with typical gas reservoirs likely below the detection limits of current observational facilities, making these systems virtually undetectable.  Thus, since a majority of PSBs in \eagle retain significant molecular gas reservoirs \citep{psbEagle}, it would follow that such systems may instead constitute a sample of `imposter PSBs', who have recently experienced a wave of feedback which has temporarily reduced their star forming efficiency.  This is in line with their findings that such PSBs contain highly disturbed gas, with only a small fraction of kinetic energy in ordered rotation \citep{psbEagle}.

In the context of this study, it follows that in our analogous sample of UBV-selected `PSB-like' galaxies defined in \S\ref{ssec:ubvSelect} (shown in \fig \ref{fig:UBV}), those dust-rich galaxies would be defined as `impostor PSBs'.  As discussed in \S\ref{ssec:HC}, such systems likely have cores containing obscured star formation, surrounded by dusty outskirts that make their photometry resemble typical PSBs. The results from this section highlight that, in addition to the traditional photometric methods used to identify PSBs, to avoid including systems with dust-obscured ongoing star formation (`imposters'), one must incorporate additional selection criteria that focuses on their current level of star formation (in addition to their photometry).

\section{Conclusion}
\label{sec:conclusion}
The aim of this paper was to holistically investigate various hypotheses that could give rise to observations of massive $(M_*>3\times10^{9}M_\odot)$ and quenched galaxies harboring significant molecular gas reservoirs such as those from \cite{rowlands, french_2015, suess_2017, Bezanson_2022, setton2025}.  The approach we took utilized the \simba hydrodynamic galaxy-formation simulations, whose novel multi-scale prescriptions have made it one of the best modern simulations for exploring galaxy evolution and baryon cycling within a cosmological context \citep{simba}.  Cosmological simulations such as this provide a powerful laboratory capable of studying the intricate interplay between a galaxy and its ISM, allowing in-depth investigations into their coevolution and the driving mechanisms connecting them.  

While many studies have individually considered hypotheses that could give rise to observations of quenched galaxies that still harbor reservoirs of molecular gas, this study is meant to jointly investigate each within a singular cosmological simulation, providing a self-consistent analysis of the relative role of each theory across space and time.  We investigated three hypotheses in total; the first of which asking about temporal effects intrinsic to the galaxy (\S\ref{ssec:HA}), while the last two deal with observational effects.  Specifically, our second hypothesis asks if the galaxy truly contains molecular hydrogen (\S\ref{ssec:HD}), while the third asks if the galaxy is truly quenched (\S\ref{ssec:HC}).

Our main results are as follows.
\begin{itemize}[nosep,leftmargin=*]
    \RaggedRight
    \vspace*{.05in}
    \item \textbf{Hypothesis A: Temporary Quiescence (\S\ref{ssec:HA})}.  
    Rejuvenation among massive galaxies is most common for molecular gas-rich galaxies ($\sim90\%$ have $\mhh/M_*>5\%$) that have not yet fully quenched ($\sim80\%$ are transitioning; \fig \ref{fig:rejuvComb}).  It is typically triggered by gas-rich mergers (\fig \ref{fig:rejuvCase}), and is therefore most important at high redshift, becoming increasingly rare toward $z=0$ (\fig \ref{fig:rejuv}).
    \vspace*{.1in}
    \item \textbf{Hypothesis B: Incorrect $\bm{\aco}$ (\S\ref{ssec:HD})}. Here, we showed that $\aco$ declines with decreasing $\ssfr\times \tH$, falling below $\acomw$ for nearly all quenched, transitioning, and weakly star-forming galaxies (\fig\ref{fig:alphaCO}).  Assuming a Milky Way-like $\aco$ in these systems will thus lead to an overestimation in $\mhh$, most severely in the most quenched systems.
    
    \vspace*{.1in}
    \item \textbf{Hypothesis C: Dust Obscuration (\S\ref{ssec:HC})}. Focusing on those galaxies that satisfy PSB-like quenching signatures (red dots in the top panel of \fig \ref{fig:UBV}), we show in \fig \ref{fig:IR_FUV} that those systems that are most dust-rich will have an IR-based SFR that accurately traces their true SFR, yet an FUV-based SFR that underestimates it.  These findings, combined with the intuition from the bottom panel of \fig\ref{fig:UBV} that the most dust-rich systems will also be the most massive and gas-rich, indicates that the UBV-selected molecular gas-rich systems from surveys like \squig\ could have star formation occurring deep within the galaxy, enshrouded by a quenched and dusty outskirt, with increasing levels of obscuration as redshifts increase (\fig\ref{fig:observer}).
    \vspace*{.1in}
\end{itemize}

By investigating all three hypotheses, we conclude that each represent a plausible mechanism for contributing to the persistence of molecular gas in seemingly quenched galaxies, with their relative importance varying across redshift and ISM conditions. For example, at high redshift (around Cosmic Noon) galaxies are both more likely to rejuvenate and host obscured star formation. A massive, gas-rich galaxy that rapidly quenched at these early times could therefore either return to star-forming activity if its molecular gas-to-stellar mass ratio is sufficiently high, or if sufficiently dusty, harbor obscured star formation that biases its photometric classification. At lower redshifts, by contrast, galaxies are more massive and easier to detect, but their $\aco$ values depart more severely from the canonical MW-like value, leading to increasingly overestimated molecular gas masses in more quenched systems.

In our discussion we put these results into the context of recent theoretical studies to show that such high-mass UBV-selected systems would actually be classified as `imposter' PSBs due to their abundant and persistent molecular gas reservoirs.  While existing observations seem to support this picture \cite{setton2025}, more results from ALMA are needed to fully understand the ISM conditions in these systems and to understand the relative role of each of our hypothesis across space and time.
\begin{acknowledgments}
\textbf{Acknowledgments}

\noindent
D.N., R.B. and J.G. are grateful for funding from the NSF via grant AST-1908137.  E.S. was additionally funded via grant NASA ADSPS23-0007.  The authors acknowledge UFIT Research Computing for providing computational resources on HiPerGator and support that have contributed to the research results reported in this publication (\url{http://it.ufl.edu/rc}).
\end{acknowledgments}

\bibliography{bib}

\appendix
\section{Merger-Driven Rejuvenation}
\label{app:merge}
\begin{figure}[H]
    \centering
    \includegraphics[width=\linewidth]{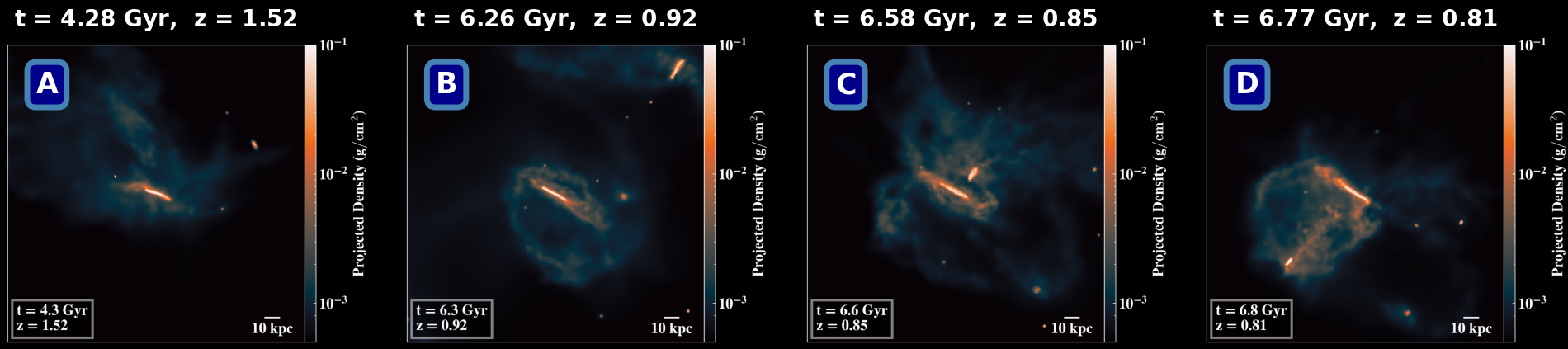}
    \caption{\textbf{A}: previously star-forming galaxy is beginning to quench; \textbf{B}: interloping galaxy enters the gravitational well of our target galaxy; \textbf{C}: interloping system falls into our target galaxy; \textbf{D}: interloping system falls through and past our target galaxy, replenishing the molecular gas in the galaxy.}
    \label{fig:rejuvABCD}
\end{figure}

\begin{figure}[H]
    \centering
    \includegraphics[width=\linewidth]{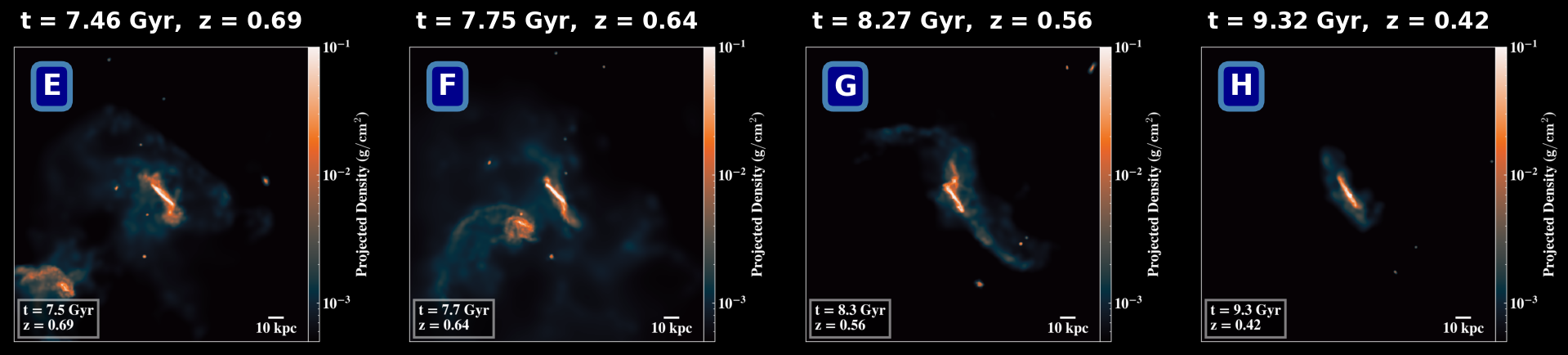}
    \caption{\textbf{E}: interloper galaxy, after first pass through, again begins to fall back into our target system; \textbf{F}: interloper galaxy falls back for its second pass through; \textbf{G}: interloper galaxy finally merges with our target galaxy; \textbf{H}: after merging, the combined systems heads towards quiescence.}
    \label{fig:rejuvEFGH}
\end{figure}

In \S\ref{ssec:HA}, while investigating whether the observed molecular gas in quenched galaxies could be due to them experiencing a temporary period of dormancy (followed by a rejuvenation), we stated that the rejuvenation event shown in \fig \ref{fig:rejuvCase} was driven by a major merging event.  In this section, we provide proof of this by visually examining the galaxy at various points in time.  To accomplish this, we pick eight points of interest during our galaxies evolution and show a projection plot of the gas density for each (shown in \fig \ref{fig:rejuvABCD}/\ref{fig:rejuvEFGH}), labeled with a letter (A-H) that corresponds to the time/redshift indicated by the same letter in the middle of \fig \ref{fig:rejuvLetters} below.

\begin{figure}[h]
    \centering
    \includegraphics[width=\linewidth]{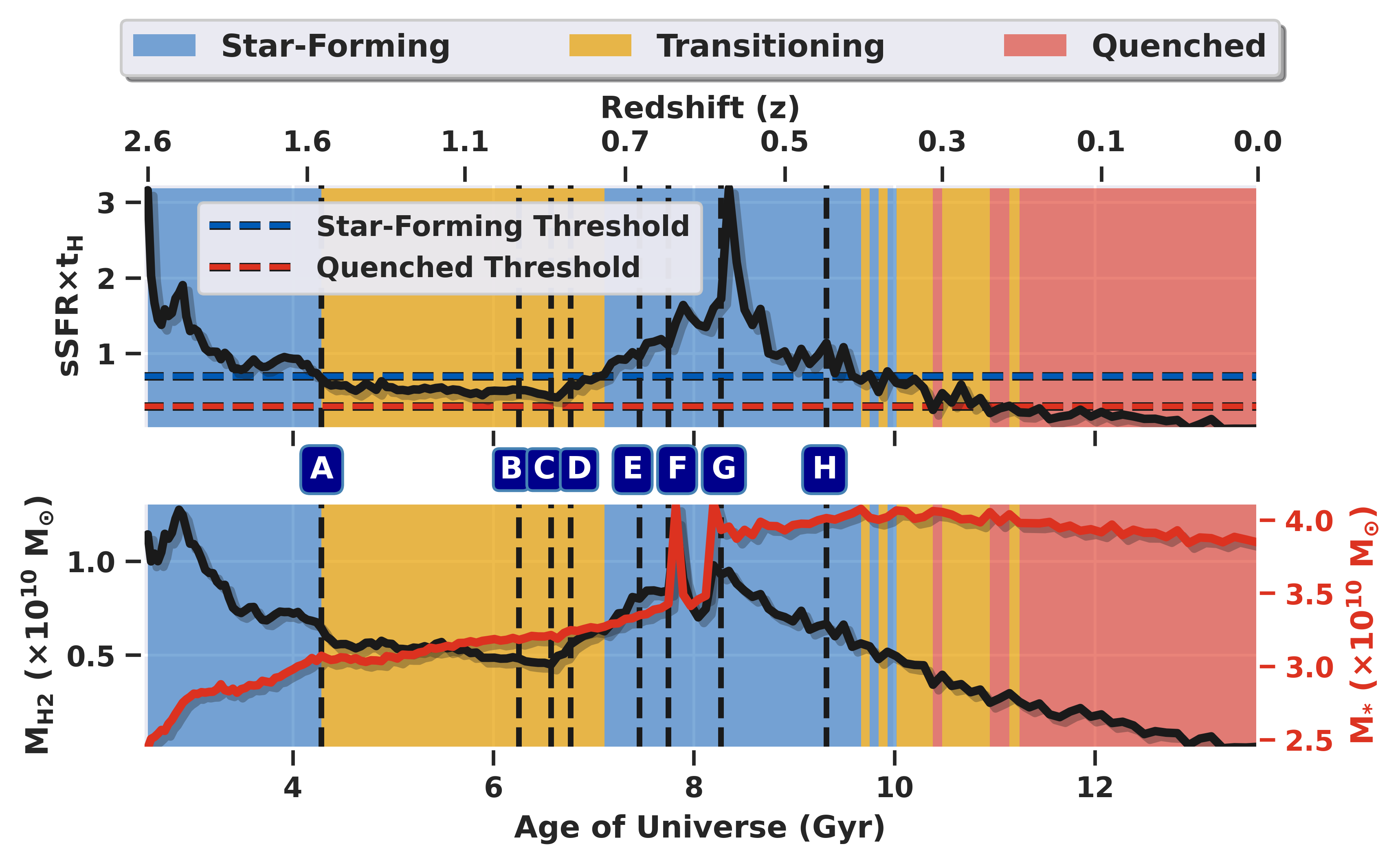}
    \caption{{\bf Example of the rejuvenating galaxy from \fig \ref{fig:rejuv} (\S\ref{ssec:HA}), with added letters (A-H) corresponding to the images in \fig\ref{fig:rejuvABCD}/\ref{fig:rejuvEFGH}.}  This galaxy was selected at $z\approx 1.5$ (`A': $\tH\sim 4.3$\,Gyr), based on it satisfying both our conditions; having $M_*>3\times 10^9M_\odot$ and $\ssfr\times\tH\leq 0.7$, with the background color set to the star-forming classification of the galaxy based on \eq \ref{eq:ssfrt1}-\ref{eq:ssfrt3}.  While the top panel shows the evolution of $\ssfr\times \tH$, the bottom panel shows the corresponding evolution of $\hh$ mass (left/black) and stellar mass (right/red), both of which show a gradual evolution at high redshift, followed by a dramatic increase at around $z\approx 0.6$ driven by a major merging event.  The drop-off after this and subsequent rise show that the merging galaxy initially passes through our main galaxy, but, now bound within its potential well, later falls back and ultimately merges.}
    \label{fig:rejuvLetters}
\end{figure}

As a reminder to the reader, in \fig \ref{fig:rejuvLetters} (as in \fig \ref{fig:rejuvCase}; \S\ref{ssec:HA}), the top panel shows the quantity we used to classify the galaxy's current state of star formation (i.e. $\ssfr\times \tH$; see \S\ref{ssec:select}), thus has two horizontal lines located at (a) the star-forming cut (\eq \ref{eq:ssfrt1}: $\ssfr\times \tH=0.7$) and (b) the quenching threshold (\eq \ref{eq:ssfrt3}: $\ssfr\times \tH=0.3$).  The background color in both panels reflects this classification, with blue regions showing where the system would be considered `star-forming', yellow for `transitioning', and red for `quenched'.  To investigate the external processes that influence this galaxies evolution, we picked eight illustrative moments across time and produced a projection plot of the gas density in our galaxy for each (\fig \ref{fig:rejuvABCD}/\ref{fig:rejuvEFGH}) via \yt \citep{yt}, an astrophysical analysis and visualization toolkit.

We begin our exploration of this galaxies evolution at a redshift of $z=1.52$ (labeled `A'; $\tH\sim 4.28$ Gyr), in which the previously star-forming galaxy has just had its SFR decrease into our transitioning phase of star formation (\eq\ref{eq:ssfrt2}).  In an isolated medium such a galaxy will continue towards quiescence, but in our cosmological simulation mergers can also occur.  We see evidence of an interloper galaxy encroaching into the territory of our target galaxy in the top-right of the image labeled `B' in \fig \ref{fig:rejuvABCD} ($\tH=6.26$ Gyr, $z=0.92$).  Moving forward in time, we can see in the same figure (\fig \ref{fig:rejuvABCD}) that this interloper passes directly through the center of our target galaxy in the images labeled `C'/`D' around $\tH=6.6$ Gyr.  At the same time in \fig \ref{fig:rejuvLetters}, we see at the lines labeled `C'/`D', that after this initial pass through, while stellar mass seems unaffected (red line), the $\hh$ mass (black line) of the galaxy begins to increase after this point $(\tH>6.6 \, \text{Gyr})$.  This behavior arises since by definition, gas is collisional while stars are not, thus in this initial pass through the gas in the two galaxies is the first to collide and collapse.

Due to the high velocity of this initial galactic collision, the interloper galaxy does not immediately coalesce with our target galaxy.  Instead, we see the interloper galaxy passes through and past our target galaxy, reaching a maximum distance from it at around 500 Myr after the initial pass-through, after which it begins to fall back into the gravitational well of our target galaxy, as shown in the image labeled `E' in \fig\ref{fig:rejuvEFGH} ($\tH=7.46$Gyr, $z=0.69$).  Around 300 Myr later is when the interloper galaxy finally falls back and begins its second pass through of our target galaxy.  This can be seen in the image labeled `F' in \fig\ref{fig:rejuvEFGH} ($\tH=7.75$Gyr, $z=0.64$).  In the same moment of time in \fig \ref{fig:rejuvLetters}, at the line labeled `F' we see this second pass through causes the galaxy tracker to briefly misidentify the galaxy, as it is not sure which galaxy is our primary during this turbulent collision.  Regardless, this merger causes a rapid rise in stellar mass at the lines labeled `F' and `G' in \fig \ref{fig:rejuvLetters} as the galaxy passes through a second time then eventually fully merges with our target galaxy, as seen in the image labeled `G' in \fig\ref{fig:rejuvEFGH} ($\tH=8.27$Gyr, $z=0.56$).

In between the points labeled `F' and `G' in \fig \ref{fig:rejuvLetters}, we see a rapid increase in the SFR of our target galaxy, as it absorbs the gas and stars from the interloping system.  After our target galaxy fully absorbs the merging system, its SFR is seen to reach a maximum value at $\tH=8.35$\,Gyr ($z=0.55$), after which it again begins to fall towards quiescence.  This galaxy will eventually deplete all of its newly replenished gas, and in our formalism is defined to be fully quenched by $\tH=11.25$ Gyr ($z=0.21$), after which it will continue as a typical elliptical galaxy towards complete quenching. 

\section{Comparing Selection Methods}
\subsection{Comparing sSFR*t to distance from SFMS}
\label{app:ssfrCompare}
As discussed in \S\ref{ssec:select}, an alternative approach to distinguishing between star-forming and quenched galaxies is based on the galaxies offset from the majority of star-forming systems on a plot of stellar mass ($\text{M}_{*}$) vs star-formation rate ($\sfr$).  Typical star-forming galaxies have a well established relationship between $\text{M}_{*}$ and $\sfr$ (known as the Star-Forming Main Sequence; SFMS; \citealt{noeske, whitaker12, sfms,Speagle14}).  This relationship arises because galaxies that lie on the main sequence are in a quasi-equilibrium (`gas-regulated') state  \citep{tacconi}, in which their star formation is limited by the amount of molecular gas available $(\mhh)$ as well as the gas' efficiency at creating stars ($t_{\text{dep}}\equiv \mhh/\sfr=$ depletion time; \citet{Saintonge2011}).
To calculate the SFMS for our simulated galaxies, we followed the formalism of \cite{sfms} to fit a second-order polynomial to the running median of log SFR at each redshift.  This was done iteratively, allowing us to fit the polynomial from \cite{sfms}, given as
\begin{equation}
\log \text{SFR}_{\text{MS}}=a+b\log(M_*/M_\odot)+c\log(M_*/M_\odot)^2\approx\log \left(\frac{\text{Molecular Gas Mass } (\mhh)}{\text{Depletion Time } (t_{\text{dep}})}\right)
\label{eq:sfms}
\end{equation}
The best fit coefficients from this model $(a,b,c)$ were found to be $(-22.59, 4.03, -0.18)$ at $z=0$, $(-20.23, 3.40, -0.13)$ at $z=1$, $(-17.39, 2.68, -0.08)$ at $z=2$, $(-14.25, 1.98, -0.04)$ at $z=3$, and $(-3.86, -0.30, 0.08)$ at $z=4$.

The resulting SFR from \eq \ref{eq:sfms} therefore represents the expected SFR of a typical star-forming galaxy on the main sequence at a given stellar mass ($M_*$), and is shown in \fig\ref{fig:sfms} (\S\ref{ssec:select}) as a solid black curve for our \simba galaxies.  Due to this definition, it follows that comparing a galaxies true SFR to its expected main sequence SFR (SFMS; \eq\ref{eq:sfms}) can provide another method of classifying a galaxy by its level of star-formation.  As a result, we continue to define how far below a galaxies true SFR is from its expected main sequence SFR using
\begin{equation}
    \Delta \text{SFMS (dex)}=\log_{10}(\text{SFR}_{\text{MS}})-\log_{10}(\text{SFR}_{\text{True}})=\log_{10}\left(\frac{\text{SFR}_{\text{MS}}}{\text{SFR}_{\text{True}}}\right)=\text{dex below SFMS}
    \label{eq:deltaSFMS}
\end{equation}

Using \eq\ref{eq:deltaSFMS}, we can now use the distance a galaxy's true SFR (\eq\ref{eq:sfr}) is below its main sequence SFR (\eq\ref{eq:sfms}) to quantify its level of quiescence.  From \cite{whitakerReview}, galaxies $1\sigma$ below the SFMS are typically 0.3 dex away, so 1 dex below would correspond to $\sim3.3\sigma$.  In this formalism, galaxies below the SFMS by $2-3\sigma$ ($0.6-0.9$ dex) would be considered `transitioning', while those further than 0.9 dex below are conservatively quenched.  To summarize, we find from \cite{whitakerReview} that galaxies can be classified by \eq\ref{eq:deltaSFMS} like
\begin{align}
    \textrm{Star Forming}:\quad & \Delta \text{SFMS}\leq 0.6~\text{dex}\label{eq:sfms1}\\
    \textrm{Transitioning}:\quad & 0.6~\text{dex}<\Delta\text{SFMS}<0.9~\text{dex}\label{eq:sfms2}\\
    \textrm{Quenched}:\quad &\Delta \text{SFMS}\geq 0.9~\text{dex}\label{eq:sfms3}
\end{align}

We show a direct comparison of this method of classifying galaxies versus our $\ssfr\times \tH$ method defined by \eq\ref{eq:ssfrt1}-\ref{eq:ssfrt3} in \fig\ref{fig:sfrBattle} below.  Here we show all $M_*>3\times 10^9M_\odot$ galaxies in our sample for $z=0-4$, with their value of $\ssfr\times\tH$ on the $x$-axis and their distance below the SFMS (\eq\ref{eq:deltaSFMS}) on the $y$-axis.  We additionally added vertical and horizontal hatched regions corresponding to the `quenched' and `star-forming' regimes of both methods in red and blue respectively, with labels describing their meaning on the top/right.  Additionally, in between each regime we have indicated via grey shading the area defined as `transitioning'.  Looking at \fig\ref{fig:sfrBattle}, we see that our fiducial quenching threshold of $\ssfr\times\tH=0.3$ roughly corresponds to 0.6 dex at $z=0$ and 0.9 dex for $z>0$. 

\begin{figure}[H]
    \centering
    \includegraphics[width=\linewidth]{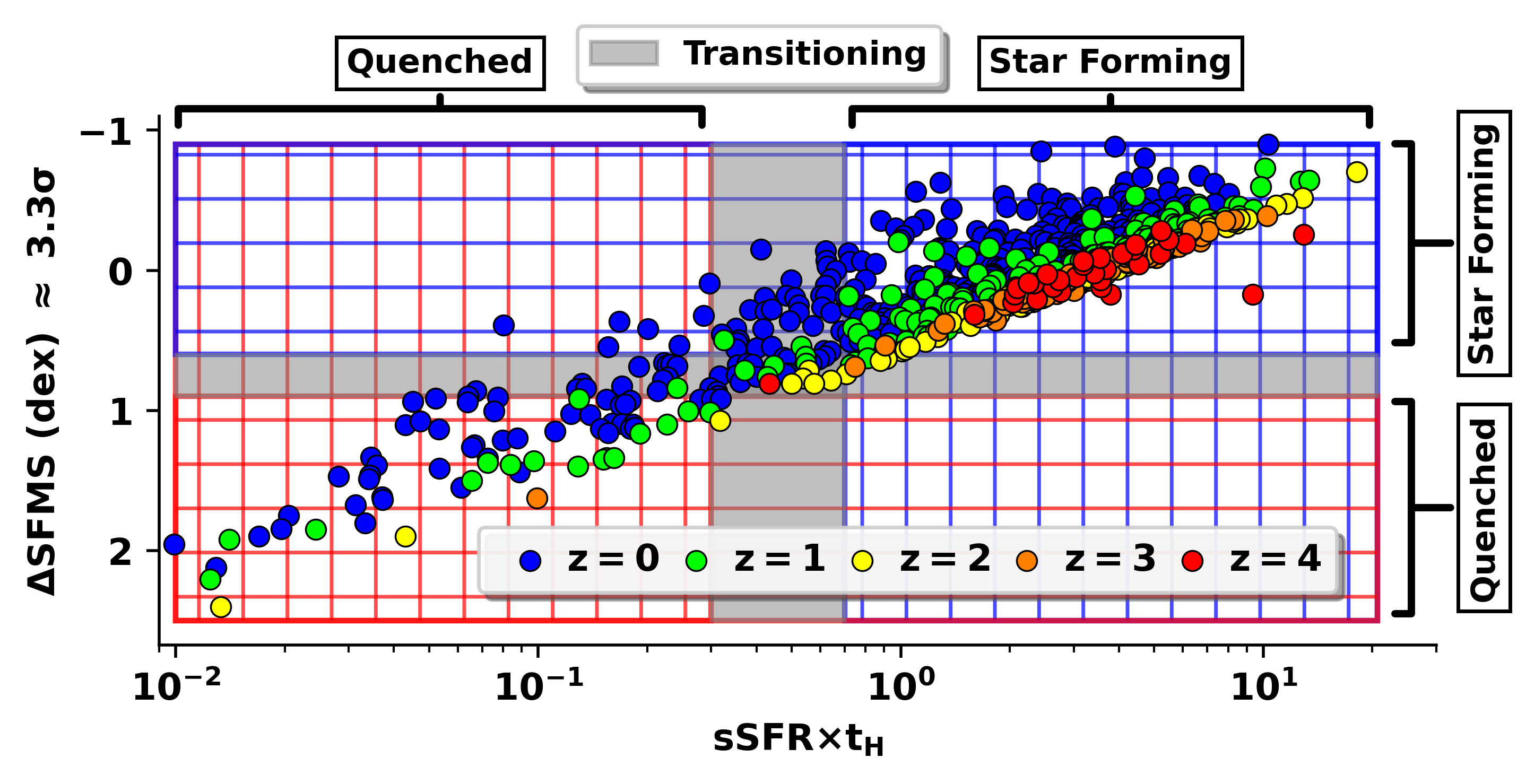}
    \caption{\textbf{Direct comparison of the galaxy classification method described in \S\ref{ssec:select} (used in \S\ref{ssec:HA}/\S\ref{ssec:HD}) vs the alternative SFMS-based classification described in this section.} All $M_*>3\times 10^9M_\odot$ galaxies are shown as dots, colored by their redshift, with their sSFR$\times$t$_\text{H}$ on the $x$-axis and their distance below the SFMS (\eq\ref{eq:deltaSFMS}) on the $y$-axis.  The quenched/star-forming regions for each definition are shown via red/blue vertical and horizontal lines, with labels on the top/right of the axii, while the transitioning regime is indicated via grey shading.}
    \label{fig:sfrBattle}
\end{figure}

\begin{figure}
    \centering
    \includegraphics[width=\linewidth]{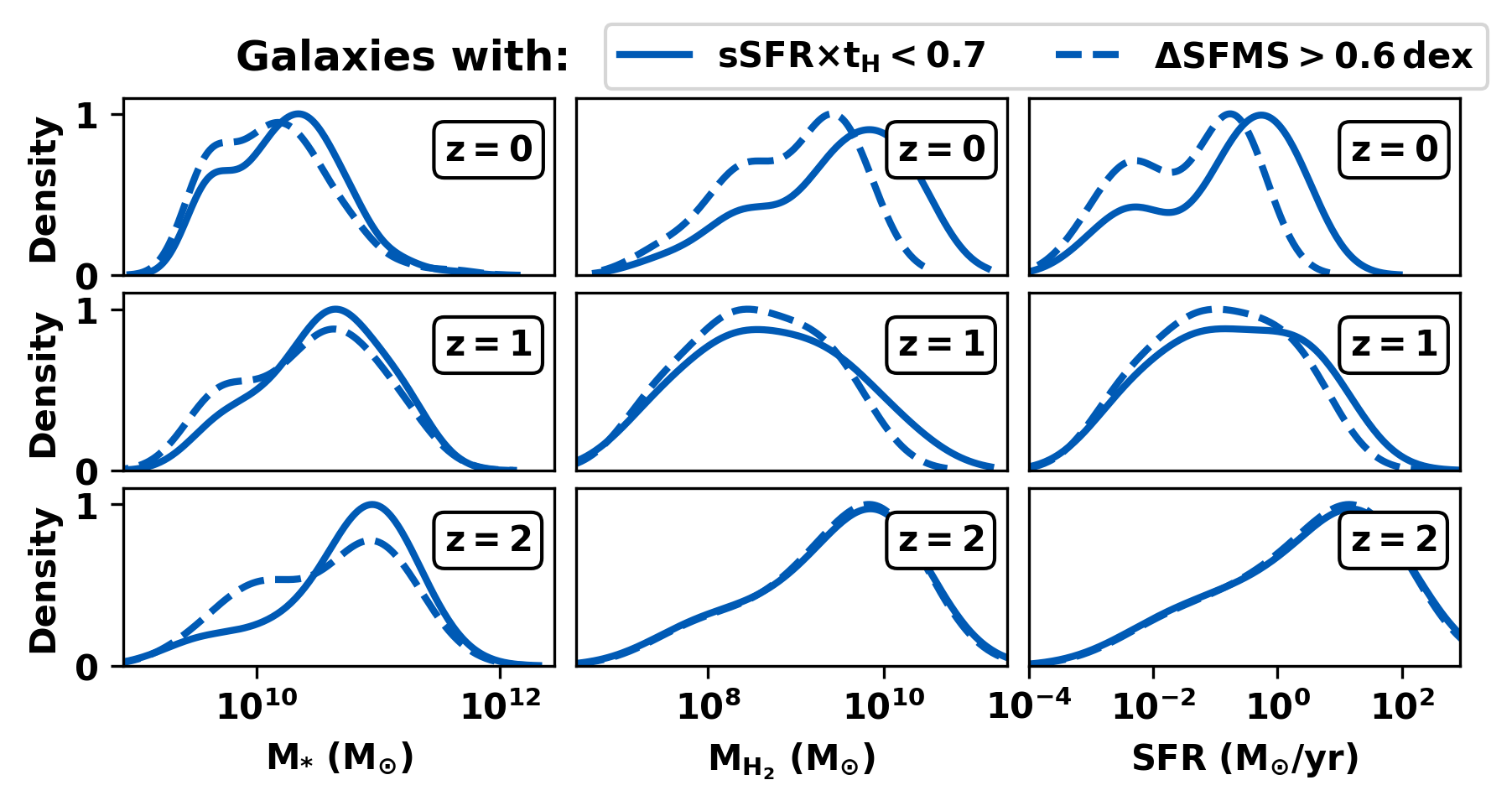}
    \caption{\textbf{Both the sSFR and $\bm{\Delta}$SFMS classification methods show good agreement at $\bm{z > 0}$, whereas substantial deviations exist at $\bm{z=0}$.}  In each panel we show the distribution of $M_*>3\times 10^9M_\odot$ galaxies identified to be `low star forming' using our sSFR-based definition ($\ssfr\times \tH<0.7$) as a solid blue line and the corresponding sample of `low star forming' galaxies identified using $\Delta$SFMS ($\Delta \text{SFMS}>0.6~\text{dex}$) as a dashed blue line.  The left column shows the distribution of stellar mass ($M_*$), the middle column shows molecular gas mass ($\mhh$), while the right-most column shows SFR, which by definition is heavily dependent on the $\mhh$.  Each row is defined by a single redshift.}
    \label{fig:sfrBattleHistograms}
\end{figure}

Since our study is broadly interested in the properties of all non star-forming galaxies, we can combine our transitioning and quenched populations into one we can call `low star forming'.  From our fiducial definitions, it follows that `low star-forming' galaxies are any with $\ssfr\times\tH< 0.7$.  Using the SFMS-based definition, it follows that `low star-forming' galaxies are any with $\Delta \text{SFMS}>0.6~\text{dex}$.  Now we can compare how the `low star-forming' galaxies used in our study compares with the analogous sample generated with the SFMS-based definition (\eq\ref{eq:sfms}).  This comparison is shown via a a table of KDE distribution plots in \fig \ref{fig:sfrBattleHistograms}.  In each panel we show the distribution of $M_*>3\times 10^9M_\odot$ galaxies identified to be `low star forming' using our sSFR-based definition as a solid blue line, and the corresponding sample of `low star forming' galaxies identified using $\Delta$SFMS as a dashed blue line.  In \fig\ref{fig:sfrBattleHistograms}, the left column shows the distribution of stellar mass ($M_*$), the middle column shows the stellar-mass weighted molecular hydrogen gas mass ($\mhh$), while the right-most column shows stellar-mass weighted star formation rate (SFR), which by definition is heavily dependent on $\mhh$ (\eq\ref{eq:sfr}).  The top row shows data for $z=0$, the middle for $z=1$, and the bottom for $z=2$.

Upon examining \fig\ref{fig:sfrBattleHistograms} we see two interesting effects.  First, the stellar mass (left-most column) from the two samples is approximately equal for all redshift, and second; while both the $\hh$ mass and SFR agree for $z=2$, they diverge at $z=0$.  It makes sense that both samples have the same stellar mass as both methods used stellar mass as fundamental inputs into their formalism (i.e. both $\ssfr$ and SFR$_{\text{MS}}$ are calculated via $M_*$).  Furthermore, despite neither method explicitly using $\mhh$, we find their resulting $\mhh$ distributions are seemingly different for $z=0$, with the $\ssfr$ method of selecting `low star forming' galaxies producing a sample with higher molecular gas mass (and SFR) than the $\Delta$SFMS method explored here.  The reasoning for this is that, as explained at the beginning of this section, galaxies on the main sequence have their star-formation regulated by their available reservoir of $\hh$, so that galaxies below the main sequence will always be relatively depleted of molecular gas.  In contrast, our $\ssfr$-based definition of `low star forming' galaxies introduces no analogous cut in molecular gas, allowing gas-rich systems with suppressed star forming efficiency to be included in our sample.  

In summary, using our sSFR-based method of identifying `low star forming' galaxies produces a sample of galaxies with higher molecular gas masses then the analogous sample selected using $\Delta$SFMS.  Thus, since the focus of our paper is on those `low star forming' galaxies that are also molecular gas-rich, we stick with the sSFR-based method.  Additionally, doing so avoids the need to iteratively fit the SFMS at every snapshot we investigate.

\subsection{UBV Selection Comparisons}
\label{app:ubv}

The sample of galaxies from \cite{suessSquiggle} comes from the Studying Quenching in Intermediate-z Galaxies: Gas, angu$\vec{L}$ar momentum, and Evolution (SQuiGG$\vec{L}$E) survey of intermediate-redshift PSBs from SDSS.  From their initial sample of $\sim$32,000 galaxies from $z=0.5-0.9$, applying their UBV cut specified above results in their final sample of $\sim$1,300 unique PSB systems.  They find the majority of their systems are massive ($M_*\gtrsim 10^{11}M_\odot$), have SFRs below the main sequence, and formed more than 10\% of their mass in a recent burst.  This is by design, as \squig\ is meant to select bright, massive, and burst-dominated galaxies, but does not directly reflect the mass of typical PSBs, as lower mass systems could exist but would not have been bright enough to be included in this catalog.  Furthermore, their final sample is also found to be associated with high H$\delta_A$ equivalent widths (implying no ongoing star formation) and low Dn4000 indices (implying a young/hot/massive stellar population).  When stacking all the spectra from their final sample, as shown as the black line in \fig \ref{fig:specStack} (grey shading indicates the 16th-84th percentiles), they find there average \squig\ spectra will have a strong Balmer break, blue slopes redwards of this break, deep Balmer absorption lines, and typically weak or absent [OII] and [OIII] emission lines.

\begin{figure}[H]
    \centering
    \includegraphics[width=\linewidth]{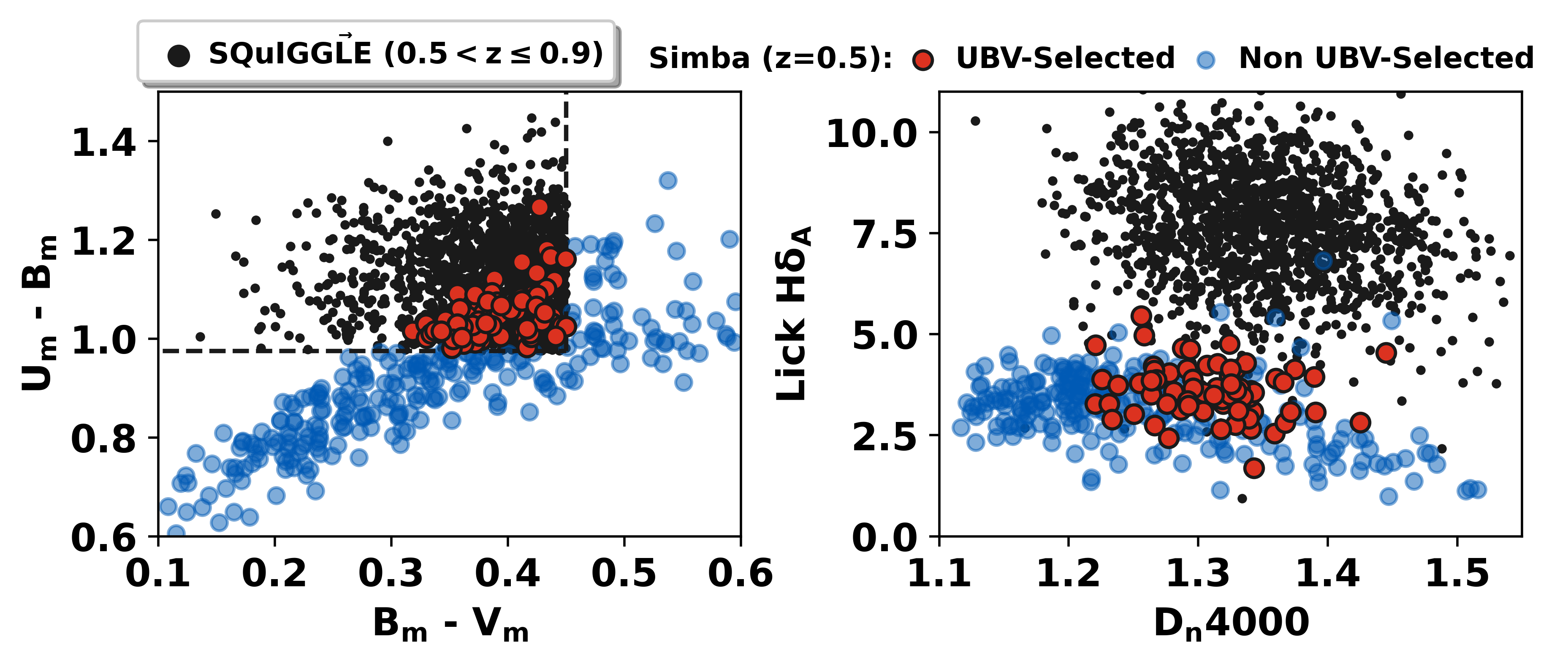}
    \caption{{\bf Same as top panel of \fig \ref{fig:UBV}, where red dots indicate those UBV-selected systems from \textbf{\textsc{SIMBA}} at $\bm{z=0.5}$, with the addition of galaxies from the SQuiGG$\bm{\vec{L}}$E survey (black points), shows how SQuiGG$\bm{\vec{L}}$E galaxies are generally redder than those UBV-selected \textbf{\textsc{SIMBA}} systems.}  As in the top panel of \fig \ref{fig:UBV}, the left-hand plot shows the galaxies in UBV space, defined by the medium-band rest-frame filters ($U_m$, $B_m$, and $V_m)$, designed to target the Balmer break and slope of the spectrum redward of the break \citep{suessSquiggle}. Right-hand plot shows the galaxies lick H$\delta_A$ index (H$\delta$ EW; which traces recent star formation) as a function of the 4000\r{A} break index (D$_{\text{n}}4000$; which probes the age of the stellar population).}
    \label{fig:ubvPlus}
\end{figure}

\begin{figure}[H]
    \centering
    \includegraphics[width=\linewidth]{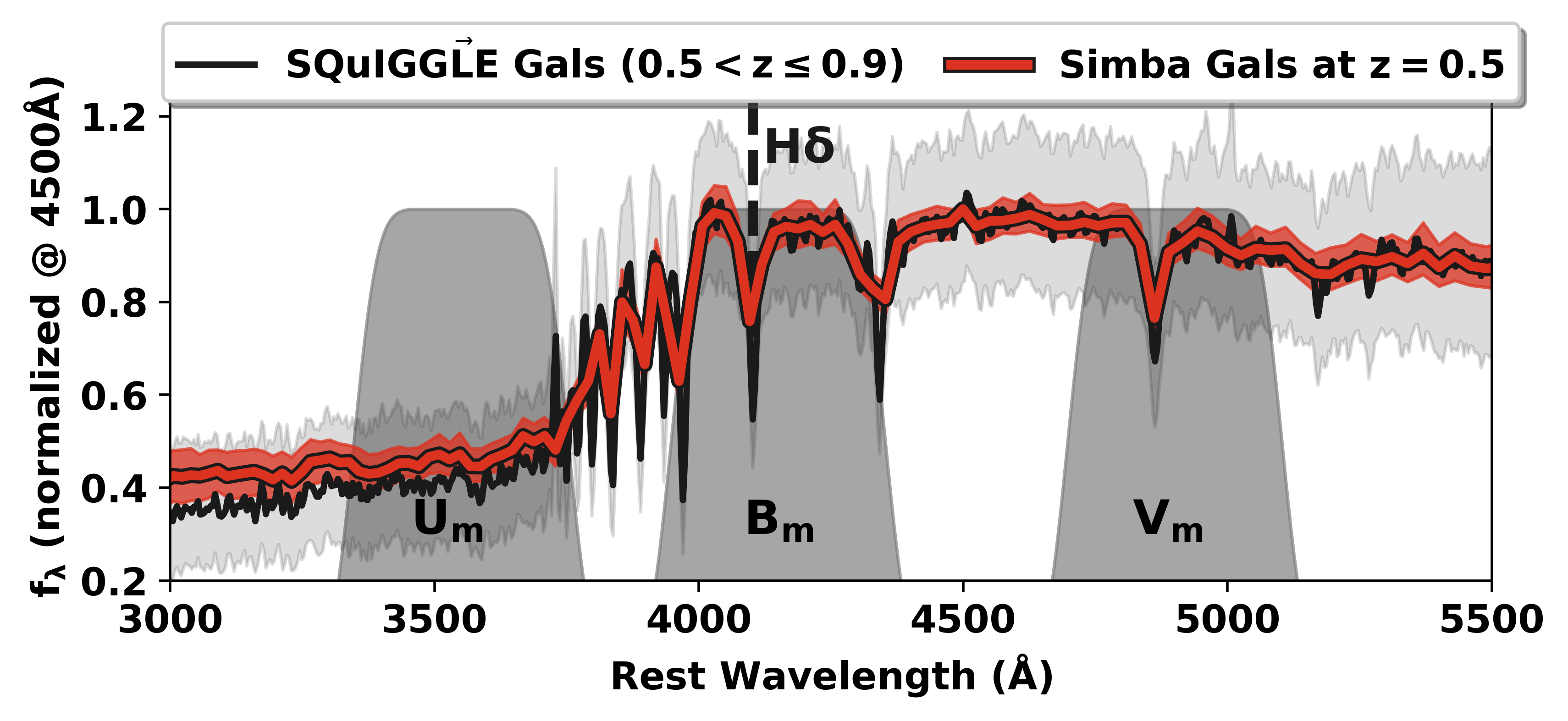}
    \caption{{\bf In general, SQuiGG$\bm{\vec{L}}$E galaxies are redder than those UBV-selected (\eq \ref{eq:UBVSelect}) galaxies from \textbf{\textsc{SIMBA}}.}  Spectral energy distributions (SEDs) for the two samples, stacked using the median, normalized at 4500\r{A}, and surrounded by a shaded region indicating the 16th-84th percentile ranges are shown above.  The black line shows the result of $\sim$1,300 stacked \squig\ galaxies from $z=0.5-0.9$, while the red line shows the combined stack of 72 UBV-selected \simba galaxies at $z=0.5$.  Additionally, we show as gray shaded regions the synthetic rest-frame $U_m$, $B_m$, and $V_m$ filters used to calculate the quantities shown in the left-panel of \fig \ref{fig:ubvPlus}.  Finally, we indicate the location of the Balmer H$\delta$ absorption line via a dashed vertical black line at $\lambda=4101.734$\r{A}.  We see here that in comparison to \squig\ galaxies, the UBV-selected systems in \simba show a weaker Balmer break and stronger UV emission, suggesting that their stellar populations are more strongly influenced by recent star formation and therefore contain a larger fraction of young, hot, and massive stars.}
    \label{fig:specStack}
\end{figure}

Now to compare their resulting sample against our analogous sample of UBV-selected galaxies (used in \S\ref{ssec:HC}), we again show the top panel of \fig \ref{fig:UBV}, but now add to the plot the analogous galaxy sample used by \citealt{suessSquiggle} (black dots).  Upon examining \fig\ref{fig:ubvPlus}, we see many differences between the two samples.  First, we see that the \squig\ galaxies (black points) have much higher/redder values of $U_m-B_m$, which implies that \squig\ galaxies have a larger Balmer break and weaker UV emission than those UBV-selected systems in \simba (red points).  Specifically, $U_m-B_m$ quantifies the contribution of young stars to the total light, so is a sensitive probe of recent star formation, which further implies that the \squig\ galaxies are also more burst dominated than those in \simba\!\!.  As a result, we see also on the right-panel of \fig\ref{fig:ubvPlus} that the fact that \simba galaxies (red points) are more dominated by recent star formation causes them to have significantly lower values of Lick H$\delta_A$ than the \squig\ galaxies (black points).

This can be confirmed visually in in \fig\ref{fig:specStack}, in which we show a combined stack (normalized at 4500\r{A}) of all $\sim$1,300 \squig\ galaxies as a solid black line, and the combined stack of our 72 UBV-selected ($z=0.5$) \simba galaxies as a solid red line, with both lines surrounded by a shaded region indicating the 16th-84th percentile range of the spectra.  Here we see that the \squig\ galaxies do have larger Balmer breaks than the corresponding \simba systems, as concluded above.  Additionally, focusing on the location of the $H\delta$ absorption line (4101\r{A}), we see that the \squig\ galaxies have significantly stronger Balmer-line absorption then the \simba galaxies.  This arises from the fact that \squig\ are more burst-dominated than \simba\!, which makes sense as by design, \simba is not meant to include bursty star formation.  The inability for \simba to produce `bursty' star formation largely stems from its subgrid treatment of the interstellar medium (ISM), which imposes a resolution-dependent temperature floor to prevent artificial fragmentation below the simulation’s mass resolution \citep{Springel03, mufasa, simba}. This effective equation of state stabilizes dense gas against collapse on unresolved scales, thereby smoothing star-formation variability in time. 

Secondly, we also see that the \squig\ galaxies (black points) also have very similar values of $B_m-V_m$ when compared to the UBV-selected \simba systems (red points), implying that the \squig\ galaxies have similar slopes in their spectra redwards/rightwards of the Balmer break than the UBV-selected systems in \simba\!\!.  This is also visually confirmed in \fig \ref{fig:specStack}, in which we see that redwards of the Balmer break (towards higher wavelengths) both spectral stacks have extraordinarily similar slopes, reflecting the similar $B_m-V_m$ distributions calculated for both samples.

\end{document}